\documentclass[english]{article}
\usepackage{geometry}
\usepackage{footnote}
\usepackage{babel}
\usepackage{float}
\usepackage{rotfloat}
\usepackage{amsthm}
\usepackage{epsfig,color,amsmath,amssymb,euscript}
\usepackage{graphicx}
\usepackage{subfig}
\usepackage{setspace}
\usepackage{url}
\usepackage{hyperref}
\usepackage[sort,comma,numbers, authoryear]{natbib}
\usepackage{authblk}
\usepackage{multirow}
\usepackage{textcomp}
\usepackage{comment}
\usepackage{xcolor}

\makeatletter

\makeatletter

\newtheorem{theorem}{Theorem}

\newtheorem{lemma}{Lemma}

\def\T{{ \mathrm{\scriptscriptstyle T} }}

\newcommand{\norm}[1]{\big\| #1 \big\|}

\def\Q{\mathbf{Q}}
\def\Z{\mathbf{Z}}
\def\R{\mathbf{R}}

\def\E{\mathbf{E}}
\def\1{\mathbf{1}}        
\def\U{\mathbf{U}}

\def\I{\mathbf{I}}
\def\D{\mathbf{D}}

\def\0{\mathbf{0}}

\def\A{\mathbf{A}}

\def\z{\mathbf{z}}
\def\B{\mathbf{B}}

\def\x{\mathbf{x}}
\def\e{\mathbf{e}}
\def\F{\mathbf{F}}
\def\M{\mathbf{M}}

\def\a{\mathbf{a}}

\def\X{\mathbf{X}}

\def \S{\mathbf{S}}
\def \I{\mathbf{I}}
\def \A{\mathbf{A}}

\def\bepsilon{\boldsymbol{\epsilon}}
\def\bSigma{\boldsymbol{\Sigma}}

\def\tr{\textnormal{tr}}

\def\log{\text{log}}

\def\f{\mathbf{f}}

\def\wt{\widetilde}

\def\cX{\mathcal{X}}

\def\Umkpre1{\hat{\U}_{\text{-}k,pre,(1)}}
\def\Umkc1{\U_{\text{-}k,(1)}}

\def\log{\textnormal{log}}

\renewcommand{\hat}{\widehat}
\renewcommand{\tilde}{\widetilde}
\usepackage{graphicx} 

\begin{document}
	\setlength{\parindent}{18pt}
	\doublespacing
	\begin{titlepage}
		
		\title{\bf Factor Strength Estimation in Vector and Matrix Time Series Factor Models}
		\author{Weilin Chen\thanks{Weilin Chen is PhD student, Department of Statistics, London School of Economics. Email: w.chen56@lse.ac.uk}}
				\author{Clifford Lam\thanks{Clifford Lam is Professor, Department of Statistics, London School of Economics. Email: C.Lam2@lse.ac.uk}}
		
		\affil{Department of Statistics, London School of Economics and Political Science}
		
		\date{}
		
		\maketitle
		
\begin{abstract}
Most factor modelling research in vector or matrix-valued time series assume all factors are pervasive/strong and leave weaker factors and their corresponding series to the noise. Weaker factors can in fact be important to a group of observed variables, for instance a sector factor in a large portfolio of stocks may only affect particular sectors, but can be important both in interpretations and predictions for those stocks. While more recent factor modelling researches do consider ``local'' factors which are weak factors with sparse corresponding factor loadings, there are real data examples in the literature where factors are weak because of weak influence on most/all observed variables, so that the corresponding factor loadings are not sparse (non-local). As a first in the literature, we propose estimators of factor strengths for both local and non-local weak factors, and prove their consistency with rates of convergence spelt out for both vector and matrix-valued time series factor models. Factor strength has an important indication in what estimation procedure of factor models to follow, as well as the estimation accuracy of various estimators \citep{Chen_Lam_2024}. Simulation results show that our estimators have good performance in recovering the true factor strengths, and an analysis on the NYC taxi traffic data indicates the existence of weak factors in the data which may not be localized.

\end{abstract}
		
		\bigskip
		\bigskip


\noindent

	\end{titlepage}
	
	\setcounter{page}{2}

\section{Introduction}
Factor modelling has become an increasingly important tool for analyzing high dimensional data across various academic fields, including finance, economics, psychology, and biology. In high dimensional time series, it is generally assumed that a small number of factors drive the dynamics of all variables, leading to significant dimension reduction. Traditional factor models primarily focus on vector time series, exploring various assumptions regarding cross-correlation and serial dependence structures \citep{BaiNg2002, Bai2003, Forni2000, ChamberlainRothschild1983, StockWatson2002, Fanetal2013, StockWatson2005, BaiNg2007, BaiNg2021, Fanetal2019, Lametal2011, LamYao2012, PanYao2008}. More recently, studies have extended their scope to include matrix factor models \citep{Wangetal2019, ChenFan2021, Heetal2022b, Yuetal2022a} and tensor factor models \citep{Chenetal2022, Hanetal2020, Hanetal2022, Heetal2022a, Barigozzi2023, Chen_Lam_2024}, incorporating emerging data in more complex matrix or tensor formats.

In factor modelling, a crucial assumption pertains to the strengths of factors. In the early studies of standard vector factor models \citep{BaiNg2002, Bai2003, StockWatson2002}, it is typically assumed that all $r$ factors are strong, commonly referred to as {\em pervasive}. Specifically, in the model
\begin{align*}
\mathbf{x}_t = \mathbf{A} \mathbf{f}_t + \bepsilon_t, \quad t = 1, \cdots , T,
\end{align*}
 where $\x_t \in \mathbb{R}^{d}$, $\A \in \mathbb{R}^{d \times r}$ and $\f_t \in \mathbb{R}^{r}$, the pervasive factor assumption implies that all $r$ eigenvalues of $\mathbf{A}^\T \mathbf{A}$ diverge proportionally to $d$, i.e., $\lambda_j(\mathbf{A}^\T \mathbf{A}) \asymp d$ for $j = 1, \cdots, r$. This results in a clear partition of the eigenvalues of the observed covariance matrix into two sets: large eigenvalues representing factor-related variation and small eigenvalues representing idiosyncratic variation. Such a clear partition is also crucial for validating the procedure to estimate the number of factors by analyzing the empirical behaviors of eigenvalues \citep{Bai2003, AhnHorenstein2013, Onatski2010}.

However, a clear separation of the eigenvalues into one set of large eigenvalues and a second set of small eigenvalues is
typically not found in practice. Empirical studies in economics and finance indicate that eigenvalues often diverge at varying rates \citep{Ross1976, TRZCINKA_1986, Freyaldenhoven2022}. In response, models introducing weak factors have been proposed for analyzing vector time series \citep{Lametal2011, LamYao2012, BaiNg2021, Freyaldenhoven2022, Onatski2012, Yoshimasa2022}. For the $j$-th column $\mathbf{a}_j$ of $\mathbf{A}$, its factor strength $\alpha_j$, ranging between 0 and 1, is defined such that
\begin{align*}
\|\a_j\|^2 \asymp d^{\alpha_j}, \ \ j = 1, \cdots, r,
\end{align*}
ensuring that
\begin{align*}
\lambda_j(\A^\T \A) \asymp d^{\alpha_j}, \ \ j = 1, \cdots, r.
\end{align*}
Thus, a strong (or pervasive) factor has $\alpha_j = 1$, while a weak factor has $\alpha_j < 1$. Theoretically, a weak factor can result from two scenarios: (i) the factor has a weak effect on all observables, or (ii) it affects only a subset of observables, referred to as a ``local'' factor by \cite{Freyaldenhoven2022}.

Building on assumptions about weak factors for vector time series, the literature has developed studies focusing on the estimation of the factor loading space and the number of factors when weak factors are present in the model \citep{Lametal2011, LamYao2012, BaiNg2021, Freyaldenhoven2022, Yoshimasa2022}. Despite these efforts, there is limited research on directly estimating factor strengths themselves. \cite{Yoshimasa2022} assume sparsity in the factor loading matrix $\mathbf{A}$ and employ techniques akin to adaptive LASSO for factor selection. They calculate the estimated factor strengths by counting the number of nonzero elements in the estimated factor loading matrix. Another study with similar sparsity assumptions, \cite{Bailey_Kapetanios_Pesaran_2021}, proposes estimating factor strengths based on the proportion of statistically significant factor loadings, but it concentrates on cases where factors are observed, while our primary emphasis is on latent factor models.

The sparsity assumptions in the above mentioned works specifically address scenarios akin to case (ii) mentioned earlier, i.e., when factors are weak due to being ``local''. This framework does not cover situations where a factor is weak because of its weak impact on all observed units. \cite{ConnorKorajczyk2022} considers such a scenario when factors are observed, and provides test-statistics for differentiating strong from weak factors. They demonstrate in an analysis of US equity returns how weak factors can have effects on some or all variables
(thus no sparsity assumption since these weak factors are not ``local''). While their test-statistics can differentiate strong from weak factors, factor strengths are not estimated in the paper. Factor strength provides important indication on how well a factor loading matrix can be estimated (see \cite{LamYao2012} and \cite{Chen_Lam_2024} for rates of convergence for factor loading matrices in the presence of weak factors).
In recent years, matrix and tensor factor models with assumptions on weak factors have also emerged \citep{Chenetal2022, Hanetal2020, Hanetal2022, Chen_Lam_2024}. However, none of these papers provide a method to estimate factor strengths. Consequently, factor strength estimation remains an important yet challenging issue, especially when we are not relying on sparsity assumptions on the factor loading matrices.


In this paper, we propose a novel method to estimate factor strengths in factor models for vector and matrix time series. Our method does not assume the factor loading matrix is sparse. Instead, we make use of covariance information and the estimated factor loading matrices to extract factor strengths directly. To the best of our knowledge, this is the first method to estimate factor strengths that can be applied in general settings when the factor loading matrices are not necessarily sparse. Moreover, it represents the first method to estimate factor strengths in matrix factor models, i.e., tensor factor models with order $K = 2$. For matrix factor models, the factor strengths on the row loading matrices and column loading matrices are estimated with specific identifiability conditions provided. Numerical experiments show that our method performs well in various settings, shedding light on future research directions in this field.

The rest of this paper is organized as follows. Section \ref{sec:vector_factor_strength} introduces our method for estimating factor strengths for vector factor models, accompanied by theoretical results. In Section \ref{sec:matrix_factor_strength}, we extend the approach to matrix factor models and provide an identifiability condition that enables the simultaneous estimation of factor strengths on both modes, with theoretical guarantees. Section \ref{sec:simulation_chapter5} presents our simulation studies, showcasing the performance of our method in various settings, with a matrix-valued NYC taxi data set analyzed in Section \ref{subsec:realdata}.  All proofs are presented in Section \ref{sec:appendix}.

\section{Estimation Method in Vector Factor Models}\label{sec:vector_factor_strength}

The models we consider are time series factor models in vector or matrix formats. We start with a vector factor model, which takes the form
\begin{align}\label{eqn:VectorFactorModel_chapter5}
\x_t = \A \f_t + \bepsilon_t, \quad t \in [T],
\end{align}
where $\x_t \in \mathbb{R}^{d}$, $\bepsilon_t \in \mathbb{R}^{d}$, $\A \in \mathbb{R}^{d \times r}$ is the factor loading matrix, and $\f_t \in \mathbb{R}^{r}$ are the latent factors. We also define, for any positive integer $m$, $[m] :={1,...,m}$. We assume $d \gg r$ and $r$ is finite, and present the following assumptions to identify the factor loadings and factor strengths.



\begin{itemize}
    \item [(V1)] (Factor strengths) {\em $\A$ is of full rank, and $\A^\T \A = \D$, where
  $\D$ is a diagonal matrix. Define the diagonal entries of $\D$ as $d_{jj}: = (\D)_{jj}$, then $d_{jj} \asymp d^{\alpha_{j}}$ for $j \in [r]$, and $0 < \alpha_{r} \leq \dots \leq \alpha_{1} \leq 1$.
  }
  \item [(V2)] (Latent factors) {\em There is $\z_{f,t}$ the same dimension as $\f_t$, such that $\f_t = \sum_{q\geq 0}a_{f,q}\z_{f,t-q}$. The time series $\{\z_{f,q}\}$ has i.i.d. elements with mean 0, variance 1 and uniformly bounded fourth order moments. The coefficients $a_{f,q}$ are so that $\sum_{q\geq 0}a_{f,q}^2 = 1$.}
\end{itemize}

Assumption (V1) defines factor strengths in the model. $\A^\T \A$ being diagonal is necessary to identify and estimate a spectrum of different factor strengths. Otherwise, different rotations could mix weak factors with stronger ones, making the factor strengths unidentifiable. This assumption is not as non-general as it appears. As an illustrative example, consider a $d\times 2$ factor loading matrix $\A = (\a_1,\a_2)$, with $\|\a_1\|^2 \asymp d^{\alpha_1}$ and $\|\a_2\|^2 \asymp d^{\alpha_2}$, $\alpha_1 > \alpha_2$. Also, write $\a_1^\T\a_2 = \|\a_1\|\cdot\|\a_2\|\cos(\theta)$. Then by the QR decomposition,
\begin{align*}
  \A = \Q\R, \; \text{ where } \; \R := \left(
                                            \begin{array}{cc}
                                              \|\a_1\| & \|\a_2\|\cos(\theta) \\
                                              0 & \|\a_2\| \\
                                            \end{array}
                                          \right),
\end{align*}
and $\Q$ is a $d\times 2$ matrix with orthogonal columns. The matrix $\R$ can in fact be written as 
\begin{align*}
  \R = \D^{1/2}\wt\R, \;\text{ where }\; \wt\R := \left(
                 \begin{array}{cc}
                   1 & \|\a_2\|\cos(\theta)/\|\a_1\| \\
                   0 & 1 \\
                 \end{array}
               \right) \;\text{ and }\; \D := \left(
                                                  \begin{array}{cc}
                                                    \|\a_1\|^2 & 0 \\
                                                    0 & \|\a_2\|^2 \\
                                                  \end{array}
                                                \right),
\end{align*}
with the entry $\|\a_2\|\cos(\theta)/\|\a_1\| = O(d^{(\alpha_2-\alpha_1)/2}) = o(1)$. Hence the common component in model (\ref{eqn:VectorFactorModel_chapter5}) can now be written as $\A\f_t = \wt\A \wt\f_t$, where $\wt\A := \Q\D^{1/2}$ and $\wt\f_t := \wt\R\f_t$. Now the new factor loading matrix indeed has $\wt\A^\T\wt\A = \D$, a diagonal matrix with $d_{jj} = \|\a_{j}\|^2 \asymp d^{\alpha_j}$ as in Assumption (V1). The new factor series $\wt\f_t$ is asymptotically the same as $\f_t$ since $\wt\R$ is asymptotically the identity matrix. If $\alpha_1=\alpha_2$, $\wt\R$ is not asymptotically the identity matrix, but we only need to trivially modify all proofs in the paper (omitted).

It's important to note that we do not impose any sparsity assumptions on the factor loading matrix $\A$, in contrast to other recent literature dealing with weak factors \citep{Yoshimasa2022, Freyaldenhoven2022, Bailey_Kapetanios_Pesaran_2021}. Consequently, a factor in our model can be weak if either (i) the factor has a weak effect on all observables, or (ii) it affects only a subset of observables. Such relaxed assumptions provide more flexibility for our approach to be used in practice. Assumption (V2) states that $\f_t$ has uncorrelated elements. Define $\F = \left[\f_{1}, \dots, \f_{T} \right]^\T$, then Assumption (V2) implies $\mathbf{E}[\frac{\F^\T \F}{T}] = \mathbf{I}_r$ and $\norm{\frac{\F^\T \F}{T}} = O_{\mathbb{P}}(1)$, facilitating the rationality of our method as described later.



To estimate the factor strengths $\alpha_j$, $j \in [r]$, note that from Assumption (V1), the factor loading matrix $\A$ can be written as $\A = \Q \D^{1/2}$, where $\Q \in \mathbb{R}^{d \times r}$ has orthogonal columns such that $\Q^\T \Q = \I_r$, and $\D \in \mathbb{R}^{r \times r}$ is a diagonal matrix defined in Assumption (V1). Since $\Q$ is orthogonal, the information about factor strengths in $\A$ is fully encapsulated in $\D$, given that $d_{jj} \asymp d^{\alpha_j}$. Consequently, we can estimate factor strengths by estimating the diagonal elements of $\D$. To achieve this, we define $\hat{\S} = \hat{\Q}^\T \hat{\bSigma}_x \hat{\Q}$, where $\hat\Q$ is an estimator of $\Q$, and $\hat{\bSigma}_x = \frac{1}{T} \sum_{t=1}^T \x_t \x_t^\T$. Then
\begin{align}
    \hat{\S} =  & \hat{\Q}^\T \Q\D^{1/2}  \left( \frac{1}{T} \sum_{t = 1}^T \f_t \f_t^\T \right) \D^{1/2}\Q^\T \hat{\Q} \notag \\ & + \hat{\Q}^\T\Q\D^{1/2} \left( \frac{1}{T} \sum_{t = 1}^T \f_t \bepsilon_t^\T \right) \hat{\Q} + \hat{\Q}^\T \left( \frac{1}{T} \sum_{t = 1}^T \bepsilon_t \f_t^\T \right) \D^{1/2}\Q^\T\hat{\Q} + \hat{\Q}^\T\left( \frac{1}{T} \sum_{t = 1}^T \bepsilon_t \bepsilon_t^\T \right) \hat{\Q}.\label{eqn:S_decom}
\end{align}
If the error terms $\bepsilon_t$ are appropriately bounded, with proper assumptions on its cross-correlation and serial dependence, the last three terms in (\ref{eqn:S_decom}) become small in comparison to the first term.
Moreover, considering that $\mathbf{E}[\f_t \f_t^\T] = \I_r$ by Assumption (V2), and assuming we have an estimator $\hat{\Q}$ that is close to $\Q$, we can make the following approximation:
\begin{align*}
    \hat{\S} \approx \hat{\Q}^\T \Q\D^{1/2}  \left( \frac{1}{T} \sum_{t = 1}^T \f_t \f_t^\T \right) \D^{1/2} \Q^\T \hat{\Q} \approx \D.
\end{align*}
In practice, the estimated $\hat\Q$ can be obtained through various approaches, depending on the model assumptions (see \cite{BaiNg2002, Lametal2011, LamYao2012, Bai_Li_2012, Bai_Liao_2016} for examples). Now, given that $\D$ is diagonal, we can directly derive the estimator for $d_{jj}$, $j \in [r]$, by using the diagonal entries of $\hat{\S}$, such that $\hat{d}_{jj} := \hat{s}_{jj}$, where $\hat{d}_{jj}$ and $\hat{s}_{jj}$ represent the $j$-th diagonal entries of $\hat\D$ and $\hat\S$, respectively. Thus, the factor strengths on $\A$ can be estimated as
\begin{align}\label{eqn:estimator_alpha_j}
    \hat{\alpha}_j = \frac{\log\left(\hat{d}_{jj}\right)}{\log(d)}, \ \ j \in [r],
\end{align}
and we can further obtain $\hat\A$ as $\hat\A = \hat\Q \hat\D^{1/2}$, where $\hat\D$ is a diagonal matrix with diagonal entries given by $\hat{d}_{jj}$.

To assess the estimator $\hat{\alpha}_j$ obtained by (\ref{eqn:estimator_alpha_j}), note that from Assumption (V1), the true factor strength $\alpha_j$ is defined as
\begin{align}\label{eqn:truestrength}
\norm{\mathbf{a}_j}^2 = C d^{\alpha_j}, \ \ j \in [r],
\end{align}
where $C$ is a constant that may vary across different $j$. Additionally, introduce the realized factor strength $\tilde{\alpha}_j$ as
\begin{align}\label{eqn:tilde_alpha}
\tilde{\alpha}_j := \frac{\log(\norm{\mathbf{a}_j}^2)}{\log(d)} = \alpha_j + \frac{\log(C)}{\log(d)}.
\end{align}
It is important to note that our estimator $\hat{\alpha}_j$ is, in fact, an estimator for $\tilde{\alpha}_j$ rather than the true $\alpha_j$, as $C$ and $\alpha_j$ are not identifiable. However, given that $C$ is a constant, when the dimension $d$ grows large, we expect $\frac{\log(C)}{\log(d)} \rightarrow 0$, leading to a negligible difference between $\tilde{\alpha}_j$ and $\alpha_j$. In the special case where $C = 1$, we always have $\tilde{\alpha}_j = \alpha_j$. In practical situations with finite samples and a moderately sized $d$, it is desirable for $C$ to be close to 1, ensuring that $\tilde{\alpha}_j$ does not significantly differ from $\alpha_j$. In such cases, $\hat{\alpha}_j$ serves as a reliable approximation to the true $\alpha_j$.



To introduce the theory for the consistency of $\hat\alpha_j$, we need the following additional assumptions:
\begin{itemize}
    \item [(V3)] (Noise series) {\em Define $\E = \left[\bepsilon_{1}, \dots, \bepsilon_{T} \right]^\T$, then $\left\|\frac{\E^\T \E}{T} \right\| = O_{\mathbb{P}}\left( 1 + \frac{d}{T} \right)$.}
    \item [(V4)] (Accuracy of the estimated $\hat\Q$) {\em The estimated factor loading $\hat\Q$ satisfies
    \begin{align}\label{eqn:rate_of_Q}
        \|\hat\Q - \Q\| =  O_{\mathbb{P}}(d^{\alpha_r - \alpha_1}).
    \end{align}}
    \item [(V5)] (Model parameters)
    {\em We assume $\alpha_r \geq \frac{\alpha_1}{2}$ and
    \begin{align}\label{eqn:rate_relationship}
        \frac{d^{1+\alpha_1 - 2 \alpha_r}}{T} = O(1).
    \end{align}}
\end{itemize}

Assumption (V3) is standard in the literature that addresses the possibly correlated noise \citep{Bai2003, BaiNg2007}. It asserts that (weak) cross-correlations and serial dependence can be allowed in the noise series, which can be inferred from more primitive conditions \citep{moon_weidner_2015, Onatski_2015}. Assumption (V4) states that the estimated $\hat\Q$ should be close to the true $\Q$, with the specified rate of convergence required. In the special case when all factors have the same strength, (\ref{eqn:rate_of_Q}) reduces to $\|\hat\Q - \Q\| = O_{\mathbb{P}}(1)$, which is naturally satisfied by any consistent estimator of $\Q$. It is important to note that depending on the method used to obtain $\hat\Q$, additional technical assumptions may be necessary to ensure the error bound of $\hat\Q$ is satisfied, although these details are not provided here. Assumption (V5) outlines the necessary relationships between the weakest and the strongest factor, as well as between $d$ and $T$. To consistently estimate $\alpha_j$, it is crucial that the weakest factor is not excessively weak compared to the strongest ones. This relationship also influences the required magnitude of $T$. Consider, for example, the scenario where the strongest factor is pervasive (i.e., $\alpha_1 = 1$). In this case, we will need $\alpha_r \geq 0.5$, and (\ref{eqn:rate_relationship}) will be automatically satisfied if the weakest factor is also pervasive (i.e., $\alpha_r = 1$). However, if $\alpha_r = 0.5$, then we will require $\frac{d}{T} = O(1)$ to fulfill the rate condition (\ref{eqn:rate_relationship}).

The following theorem shows the consistency of estimated factor strengths $\hat{\alpha}_j$ obtained by (\ref{eqn:estimator_alpha_j}).

\begin{theorem}\label{thm:vector_consistency}
    Under Assumptions (V1)-(V5), if the constant $C$ defined in (\ref{eqn:truestrength}) is unknown, we have
    \begin{align*}
        |\hat\alpha_j - \alpha_j| = O_{\mathbb{P}}\left(\frac{1}{\log(d)}\right) , \ \ j \in [r].
    \end{align*}
    \end{theorem}

Theorem \ref{thm:vector_consistency} asserts that $\hat{\alpha}_j$ converges to the true factor strength $\alpha_j$ with a rate of $1/\log(d)$ when we do not know the constant $C$ defined in (\ref{eqn:truestrength}). Indeed, this rate is optimal, as we have demonstrated that $\hat{\alpha}_j$ is an estimator for the realized factor strength $\tilde{\alpha}_j$ defined in (\ref{eqn:tilde_alpha}) when $C \neq 1$. Consequently, $\tilde{\alpha}_j$ converges to the true $\alpha_j$ with a rate of $1/\log(d)$, and the rate of $|\hat{\alpha}_j - \alpha_j|$ cannot surpass this bound. Theorem \ref{thm:vector_consistency} highlights that, with proper assumptions, we can achieve this optimal rate.

Nevertheless, if we assume $C = 1$, then $\tilde{\alpha}_j = \alpha_j$, making the factor strength $\alpha_j$ exactly identifiable. In such case, we can achieve a better rate of convergence for $|\hat\alpha_j - \alpha_j|$. To accomplish this, we need the following Assumption (V4') and (V5').

\begin{itemize}
\item [(V4')] {\em The estimated factor loading $\hat\Q$ satisfies
    $
        \|\hat\Q - \Q\| =  o_{\mathbb{P}}(d^{\alpha_r - \alpha_1}).
    $
    }
    \item [(V5')] {\em
    We assume $\alpha_r > \frac{\alpha_1}{2}$ and
    $
        \frac{d^{1+\alpha_1 - 2 \alpha_r}}{T} = o(1).
    $
    }
\end{itemize}

Assumptions (V4') and (V5') are parallel to Assumptions (V4) and (V5), respectively. The slightly more restrictive rate conditions are necessary for the proof of the following theorem.
\begin{theorem}\label{thm:vector_improved_rate}
    Under Assumption (V1), (V2), (V3), (V4) and (V5'), if the constant $C$ defined in (\ref{eqn:truestrength}) equals 1, then we have, for any $j \in [r-1]$ satisfying $\alpha_j > \alpha_r$,
    \begin{align}\label{eqn:rate1}
        |\hat\alpha_j - \alpha_j| =  O_{\mathbb{P}}\left(\frac{c_j + d^{\alpha_r - \alpha_j}}{\log(d)}  \right),
    \end{align}
    where
    \begin{align*}
        c_j :=  (d^{\frac{\alpha_1}{2} - \alpha_j})(1 + d^{1/2}T^{-\frac{1}{2}}) + d^{- \alpha_j}(1 + d T^{-1}) = o(1).
    \end{align*}
    Furthermore, if Assumption (V4') is satisfied, then for any $j \in [r]$,
    \begin{align}\label{eqn:rate2}
        |\hat\alpha_j - \alpha_j| =  O_{\mathbb{P}}\left(\frac{c_j}{\log(d)}  \right) + o_{\mathbb{P}}\left(\frac{d^{\alpha_r - \alpha_j}}{\log(d)} \right).
    \end{align}
\end{theorem}

Theorem \ref{thm:vector_improved_rate} presents the improved rate of convergence for the estimated factor strengths when they are exactly identifiable. Equation (\ref{eqn:rate1}) indicates that when $\alpha_r$ is not too small and $T$ is large enough to satisfy (V5'), all factor strengths except for the weakest ones can be estimated at a rate faster than $1/\log(d)$, assuming some factors are stronger than others. Moreover, stronger factors can achieve faster rates, and the rate increases as $T$ or $d$ grows larger.

Furthermore, (\ref{eqn:rate2}) states that if $\Q$ is accurately estimated such that (V4') is also satisfied, then all factor strengths, including the weakest ones, can be estimated at an even faster rate. Note that if all factors have the same strengths, then (V4') is automatically satisfied for any consistent estimator of $\Q$, and the rate (\ref{eqn:rate2}) directly applies.

\section{Extension to matrix factor models}\label{sec:matrix_factor_strength}

In Section \ref{sec:vector_factor_strength}, we discuss our method to estimate factor strengths in a vector factor model. The similar approach can be extended to a matrix factor model, which is developed for analyzing time series observations recorded in matrix form \citep{Wangetal2019, ChenFan2021, Heetal2021, Yuetal2022a}.
Consider the matrix factor model:
\begin{align}\label{eqn:MatrixFactorModel_chapter5}
\X_t = \A_1 \F_t \A_2^\T + \E_t, \quad t \in [T],
\end{align}
where $\X_t \in \mathbf{R}^{d_1 \times d_2}$, $\E_t \in \mathbf{R}^{d_1 \times d_2}$, $\F_t \in \mathbf{R}^{r_1 \times r_2}$, and $\A_k \in \mathbf{R}^{d_k \times r_k}$ for $k = 1, 2$.
The following assumptions for matrix factor models are direct extensions of Assumptions (V1) and (V2) for vector factor models:
\begin{itemize}
    \item [(M1)] (Factor strengths) {\em For $k = 1,2$, $\A_k$ is of full rank, and $\A_k^\T \A_k = \D_k$, where
  $\D_k$ is a diagonal matrix. Define the diagonal entries of $\D_k$ as $d_{k,jj}: = (\D_k)_{jj}$, then $d_{k,jj} \asymp d^{\alpha_{k,j}}$ for $j \in [r_k]$, and $0 < \alpha_{k, r_k} \leq \dots \leq \alpha_{k,1} \leq 1$.
  }
  \item [(M2)] (Latent factors) {\em There is $\Z_{f,t}$ the same dimension as $\F_t$, such that $\F_t = \sum_{q\geq 0}a_{f,q}\Z_{f,t-q}$. The time series $\{\Z_{f,q}\}$ has i.i.d. elements with mean 0, variance 1 and uniformly bounded fourth order moments. The coefficients $a_{f,q}$ are so that $\sum_{q\geq 0}a_{f,q}^2 = 1$.}
\end{itemize}

Assumption (M2) is parallel to the assumptions made for the factor series in \cite{Chen_Lam_2024} when $K = 2$. Assumption (M1) fixes the concept of factor strength similar to Assumption (V1) in Section \ref{sec:vector_factor_strength}.
With Assumption (M1), we can write $\A_1 = \Q_1 \D_1^{1/2}$ and $\A_2 = \Q_2 \D_2^{1/2}$, where $\Q_k \in \mathbf{R}^{d_k \times r_k}$ has orthogonal columns for $k = 1, 2$. Then (\ref{eqn:MatrixFactorModel_chapter5}) can be written as
\begin{align*}
    \X_t = \Q_1 \D_1^{1/2} \F_t \D_2^{1/2} \Q_2^\T + \E_t, \ \ t \in [T].
\end{align*}

To estimate the factor strengths on $\A_1$, similar to the vector case, we can create $\hat{\S}_1 = \hat{\Q}_1^\T \hat{\bSigma}_{1x} \hat{\Q}_1$, where $\hat\Q_1$ is an estimator of $\Q_1$, and $\hat{\bSigma}_{1x} = \frac{1}{T} \sum_{t=1}^T \X_t \X_t^\T$. Then
\begin{align}
    \hat{\S}_1 =  & \hat{\Q}_1^\T \Q_1\D_1^{1/2}  \left( \frac{1}{T} \sum_{t = 1}^T \F_t \D_2 \F_t^\T \right) \D_1^{1/2}\Q_1^\T \hat{\Q}_1 \notag \\ & + \hat{\Q}_1^\T\Q_1\D_1^{1/2} \left( \frac{1}{T} \sum_{t = 1}^T \F_t \D_2^{1/2} \Q_2^\T \E_t^\T \right) \hat{\Q}_1 + \hat{\Q}_1^\T \left( \frac{1}{T} \sum_{t = 1}^T \E_t \Q_2 \D_2^{1/2} \F_t^\T \right) \D_1^{1/2}\Q_1^\T\hat{\Q} + \hat{\Q}_1^\T\left( \frac{1}{T} \sum_{t = 1}^T \E_t \E_t^\T \right) \hat{\Q}_1.\label{eqn:S1_decom}
\end{align}
The last three terms in (\ref{eqn:S1_decom}) will become small compared to the first term if the error terms $\E_t$ are small with proper assumptions on its cross-correlation and serial dependence. For matrix factor models, literature has been developed to obtain $\hat\Q_1$ using different approaches under various model assumptions (see \cite{Wangetal2019, Chenetal2022, ChenFan2021, Chen_Lam_2024, Heetal2021, Heetal2022a} for examples). If $\hat\Q_1$ is close to $\Q_1$, then
\begin{align}
    \hat{\S}_1
    &\approx \hat{\Q}_1^\T \Q_1 \D_1^{1/2} \left(\frac{1}{T} \sum_{t=1}^{T}\F_t \D_2 \F_t^\T \right) \D_1^{1/2} \Q_1^\T \hat{\Q}_1 \notag \\
    &\approx \D_1^{1/2}\left(\frac{1}{T} \sum_{t=1}^{T}\F_t \D_2 \F_t^\T \right) \D_1^{1/2} \notag\\
    &\approx \D_1^{1/2}\tr(\D_2) \D_1^{1/2} \notag\\
    &= \tr(\D_2) \D_1. \label{eqn:S1_final}
\end{align}
If $\D_2$ is known, or we have an estimate for it, we can then estimate the diagonal entries of $\D_1$ by using the diagonal entries of $\hat{\S}_1 / \tr(\D_2)$. With $\hat\Q_2$ an estimator of $\Q_2$ and if they are close, parallel arguments (by swapping the index 1 with 2 and vice versa) show
\begin{align}
    \hat{\S}_2
    &\approx \tr(\D_1) \D_2. \label{eqn:S2_final}
\end{align}
Thus, if $\D_1$ is known or if we have an estimate of it, then we can estimate the diagonal entries of $\D_2$ by using the diagonal entries of $\hat{\S}_2 / \tr(\D_1)$.

However, in most practical scenarios, neither $\D_1$ nor $\D_2$ is known. Consequently, we aim to estimate both $\D_1$ and $\D_2$ simultaneously from (\ref{eqn:S1_final}) and (\ref{eqn:S2_final}). In such situations, it is crucial to note that due to matrix multiplication, the factor strengths on $\A_1$ and $\A_2$ are not identifiable in matrix factor models. This lack of identifiability  is reflected in the relationships derived from (\ref{eqn:S1_final}) and (\ref{eqn:S2_final}):
\begin{align}\label{eqn:traceS1S2}
\tr(\hat{\S}_1) \approx \tr(\D_1) \tr(\D_2) \approx \tr(\hat{\S}_2).
\end{align}
Therefore, to estimate the factor strengths on $\D_1$ and $\D_2$ simultaneously, it is necessary to define the identifiability condition as:
\begin{itemize}
\item[(IC)] (Factor strength identifiability)
\begin{align}\label{eqn:new_identifiability_condition}
\frac{\tr(\D_1)}{r_1 d_1} = \frac{\tr(\D_2)}{r_2 d_2}.
\end{align}
\end{itemize}
Note that the identifiability condition is not unique. However, we choose (\ref{eqn:new_identifiability_condition}) as it is convenient for interpretations. The intuition behind (\ref{eqn:new_identifiability_condition}) is that, in general, larger factor strengths will be ``assigned" to larger dimensions. For instance, consider $r_1 = r_2 = 1$ and $\alpha_{1,1} = \alpha_{2,1} = 1$. In this case, $\tr(\A_1^\T \A_1) = \tr(\D_1)\approx r_1d_1$ and $\tr(\A_2^\T\A_2) = \tr(\D_2)\approx r_2d_2$ (each up to multiplication of an unknown constant). Consequently, by (\ref{eqn:new_identifiability_condition}), the estimated factor strengths can recover the true ones if we know one of it, i.e., $\hat\alpha_{1,1} \approx \hat\alpha_{2,1} \approx 1$. On the other hand, if $\A_1$ and $\A_2$ have the exact same dimensions ($r_1 = r_2$ and $d_1 = d_2$), they will be ``assigned" the same factor strengths under (IC), as the factor strengths on them are completely symmetric and indistinguishable from each other.

With identifiability condition (\ref{eqn:new_identifiability_condition}), together with (\ref{eqn:traceS1S2}), we can allocate the proper factor strengths on $\D_1$ and $\D_2$ accordingly. For more accuracy and consistency in calculations, we can use the average of $\tr(\hat{\S}_1)$ and $\tr(\hat{\S}_2)$ as an estimate of $\tr(\D_1) \tr(\D_2)$ and solve for $\tr(\D_1)$ and $\tr(\D_2)$, respectively. This leads to the following approximations:
\begin{align}\label{eqn:tr1_approx}
\tr(\D_1) \approx \left(\frac{\tr(\hat{\S}_1 + \hat\S_2)}{2} \cdot \frac{r_1 d_1}{r_2 d_2}\right)^{1/2},
\end{align}
and
\begin{align}\label{eqn:tr2_approx}
\tr(\D_2) \approx \left(\frac{\tr(\hat{\S}_1 + \hat\S_2)}{2} \cdot \frac{r_2 d_2}{r_1 d_1}\right)^{1/2}.
\end{align}

By substituting (\ref{eqn:tr2_approx}) and (\ref{eqn:tr1_approx}) back into (\ref{eqn:S1_final}) and (\ref{eqn:S2_final}), we can estimate the diagonal entries of $\D_1$ and $\D_2$ by taking the corresponding diagonal entries in $\hat{\S}_1$ and $\hat{\S}_2$, respectively, normalized to specific magnitudes. This leads to:
\begin{align*}
\hat{d}_{1,jj} := \frac{\hat{s}_{1,{jj}}}{\left(\frac{\tr(\hat{\S}_1 + \hat\S_2)}{2} \cdot \frac{r_2 d_2}{r_1 d_1}\right)^{1/2}}, \ \ j \in [r_1],
\end{align*}
where $\hat{d}_{1,jj}$ and $\hat{s}_{1,{jj}}$ are the $j$-th diagonal entries of $\hat\D_1$ and $\hat\S_1$, respectively, and
\begin{align*}
\hat{d}_{2,jj} := \frac{\hat{s}_{2,{jj}}}{\left(\frac{\tr(\hat{\S}_1 + \hat\S_2)}{2} \cdot \frac{r_1 d_1}{r_2 d_2}\right)^{1/2}}, \ \ j \in [r_2],
\end{align*}
where $\hat{d}_{2,jj}$ and $\hat{s}_{2,{jj}}$ are the $j$-th diagonal entries of $\hat\D_2$ and $\hat\S_2$, respectively. Finally, the factor strengths on $\A_1$ and $\A_2$ can be estimated as:
\begin{align*}
    \hat{\alpha}_{1,j} = \frac{\log\left(\hat{d}_{1,jj}\right)}{\log(d_1)}, \ \ j \in [r_1],
\end{align*}
and
\begin{align*}
    \hat{\alpha}_{2,j} = \frac{\log\left(\hat{d}_{2,jj}\right)}{\log(d_2)}, \ \ j \in [r_2],
\end{align*}
and we can further obtain $\hat\A_k = \hat\Q_k \hat\D_k^{1/2}$, where $\hat\D_k$ is a diagonal matrix with diagonal entries given by $\hat{d}_{k,jj}$, for $k = 1,2$.

{

To introduce the theoretical guarantee for $\hat{\alpha}_{1,j}$ and $\hat{\alpha}_{2,j}$, we similarly define the following assumptions for the matrix factor models, as an extension of Assumptions (V3) to (V5) for vector factor models:
\begin{itemize}
    \item [(M3)] (Noise series) {\em $\left\|\frac{1}{d_2T}\sum_{t = 1}^T \frac{\E_t \E_t^\T}{T} \right\| = O_{\mathbb{P}}\left( 1 + \frac{d_1}{d_2T} \right)$ and $\left\|\frac{1}{d_1T}\sum_{t = 1}^T \frac{\E_t^\T \E_t}{T} \right\| = O_{\mathbb{P}}\left( 1 + \frac{d_2}{d_1T} \right)$.}
    \item [(M4)] (Accuracy of the estimated $\hat\Q_k$) {\em For $k = 1,2$, the estimated factor loading $\hat\Q_k$ satisfies
    \begin{align*}
        \|\hat\Q_k - \Q_k\| =  o_{\mathbb{P}}(d^{\alpha_{k,r_k} - \alpha_{k,1}}).
    \end{align*}}
    \item [(M5)] (Model parameters)
    {\em For $k = 1, 2$, we assume $\alpha_{k,r_k} \geq \frac{\alpha_{k,1}}{2}$, $\alpha_{k,1} \geq 0.5$. Furthermore
    \begin{align*}
        \frac{d_1}{d_2^{\alpha_{2,1}} T} = o(1), \ \ \
        \frac{d_2}{d_1^{\alpha_{1,1}} T} = o(1), \ \ \
        \frac{d_1^{1-\alpha_{1,1}}}{d_2^{\alpha_{2,r_2}}} = o(1), \ \ \ \frac{d_2^{1-\alpha_{2,1}}}{d_1^{\alpha_{1,r_1}}} = o(1).
    \end{align*}}
\end{itemize}

Assumption (M3) extends Assumption (V3) from the vector model to the matrix model, allowing for (weak) cross-correlations among fibers and serial dependence in the noise series. Specifically, we can express $\frac{1}{d_2T}\sum_{t = 1}^T \frac{\E_t \E_t^\T}{T}$ as $\frac{1}{d_2T}\sum_{t = 1}^T \sum_{i = 1}^{d_2} \e_{t,i} \e_{t,i}^T$, where the $\e_{t,i}$'s represent the columns of $\E_t$. Then Assumption (M3) will be satisfied by applying random matrix theory if the correlations among columns of $\E_t$, rows of $\E_t$, and serial dependence of $\E_t$ are not too strong \citep{AhnHorenstein2013, BaiYin1993}. Assumption (M4) states that the estimated $\hat\Q_k$ should be close to the true $\Q_k$. Consider a common scenario that $d_1 \asymp d_2 \asymp T$, and the strongest factors for both modes are pervasive, i.e., $\alpha_{1,1}= \alpha_{2,2} = 1$.  Then the rate requirement for Assumption (M4) can be satisfied by the projection estimator of \cite{Chen_Lam_2024} when $\alpha_{k,r_k} \geq 0.5$ (representing a very weak factor). In the special case when all factors have the same strength, Assumption (M4) is naturally satisfied by any consistent estimator of $\Q_k$. Assumption (M5) delineates the requisite relationships between the weakest and strongest factors of each mode, as well as among $d_1$, $d_2$, and $T$. These conditions are relatively mild, as $\alpha_{k,1} \geq 0.5$ denotes a very weak strongest factor for each mode.
In a typical scenario where $d_1 \asymp d_2 \asymp T$ and $\alpha_{1,1}= \alpha_{2,2} = 1$, Assumption (M5) will be satisfied as long as $\alpha_{k,r_k} \geq 0.5$.

Similar to the vector factor model, the true factor strength $\alpha_{k,j}$ for a factor under a matrix factor model is defined as
\begin{align}\label{eqn:truestrength_matrix}
\norm{\mathbf{a}_{k,j}}^2 = C d_k^{\alpha_{k,j}}, \ \ j \in [r_k], \ \ k \in [1,2],
\end{align}
where $C$ is a constant that may vary across different $k,j$. Additionally, define the realized factor strength $\tilde{\alpha}_{k,j}$ as
\begin{align*}
\tilde{\alpha}_{k,j} := \frac{\log(\norm{\mathbf{a}_{k,j}}^2)}{\log(d_k)} = \alpha_{k,j} + \frac{\log(C)}{\log(d_k)},  \ \ j \in [r_k], \ \ k \in [1,2].
\end{align*}
Similar to the vector case, our estimator $\hat{\alpha}_{k,j}$ is an estimator for $\tilde{\alpha}_{k,j}$ rather than the true $\alpha_{k,j}$.
When the constant $C \neq 1$, the convergence rate of $\hat{\alpha}_{k,j}$ towards the true $\alpha_{k,j}$ cannot be faster than $1/\log(d_k)$. However, if $C=1$, then the rate of convergence can be much faster. The following theorem shows the consistency of estimated factor strengths $\hat{\alpha}_{k,j}$ for matrix factor models by specifying the rates under different scenarios.

\begin{theorem}\label{thm:matrix_consistency}
     Under Assumptions (M1)-(M5), and assuming the identifiability condition (\ref{eqn:new_identifiability_condition}) holds. For each $j \in [r_k]$, $k = 1,2$, if the constant $C$ defined in (\ref{eqn:truestrength_matrix}) is unknown, then we have
     \begin{align*}
        |\hat\alpha_{k,j} - \alpha_{k,j}| = O_{\mathbb{P}}\left(\frac{1}{\log(d_k)}\right).
    \end{align*}
    Furthermore,  if $\alpha_{k,r_k} > \frac{\alpha_{k,1}}{2}$ and the constant $C$ defined in (\ref{eqn:truestrength_matrix}) equals 1, then for $j \in [r_k]$, $k =1,2$,
    \begin{align}\label{eqn:improved_rate_matrix}
        |\hat\alpha_{k,j} - \alpha_{k,j}| =  O_{\mathbb{P}}\left(\frac{c_{k,j}}{\log(d_k)}  \right) + o_{\mathbb{P}}\left(\frac{d^{\alpha_{k,r_k} - \alpha_{k,j}}}{\log(d_k)} \right),
    \end{align}
    where
    \begin{align*}
        c_{1,j} :=  d_1^{\frac{\alpha_{1,1}}{2} - \alpha_{1,j}} d_2^{\frac{1-\alpha_{2,1}}{2}} \left(1 + \frac{d_1}{d_2T}\right)^{1/2} + d_1^{-\alpha_{1,j}} d_2^{1-\alpha_{2,1}}\left(1 + \frac{d_1}{d_2T}\right) = o(1), \\
        c_{2,j} :=  d_2^{\frac{\alpha_{2,1}}{2} - \alpha_{2,j}} d_1^{\frac{1-\alpha_{1,1}}{2}} \left(1+ \frac{d_2}{d_1T}\right)^{1/2} + d_2^{-\alpha_{2,j}} d_1^{1--\alpha_{1,1}}\left(1 + \frac{d_2}{d_1T}\right) = o(1).
    \end{align*}
\end{theorem}

Theorem \ref{thm:matrix_consistency} extends the results of Theorem \ref{thm:vector_consistency} and Theorem \ref{thm:vector_improved_rate} from vector time series to matrix time series. Specifically, in general scenarios when $C \neq 1$, the optimal rate of $1/\log(d_k)$ is achieved by our estimated factor strengths. In the special case when $C = 1$ such that the factor strengths are exactly identifiable, then we can obtain an improved rate of convergence as outlined by (\ref{eqn:improved_rate_matrix}).
The improved rate (\ref{eqn:improved_rate_matrix}) for matrix factor models can be compared to the rate (\ref{eqn:rate2}) for vector factor models in Theorem \ref{thm:vector_improved_rate}. If the strongest factor for mode-2 is pervasive, i.e., $\alpha_{2,1} = 1$, then for estimating the factor strengths in $\A_1$, we have
\begin{align*}
    c_{1,j} = d_1^{\frac{\alpha_{1,1}}{2} - \alpha_{1,j}} \left(1 + \frac{d_1}{d_2T}\right)^{1/2} + d_1^{-\alpha_{1,j}} \left(1 + \frac{d_1}{d_2T}\right),
\end{align*}
which is faster than the rate $c_j$ in Theorem \ref{thm:vector_improved_rate} when $d_1 \succ d_2T$, and  at the same rate as $c_j$ when $d_1 \preceq d_2T$. Thus, with the matrix factor model, we can potentially obtain a more accurate estimator of the factor strengths.
}

\section{Simulation Experiments}\label{sec:simulation_chapter5}

In this section, we conduct simulation experiments to test the performances of our proposed method to estimate factor strengths in vector and matrix factor models.

\subsection{Simulation settings}\label{subsec:simulation_setting_chapter5}
For generating our data, we use model (\ref{eqn:VectorFactorModel_chapter5}) for vector time series, and (\ref{eqn:MatrixFactorModel_chapter5}) for matrix time series.
For vector time series, the factor loading matrix $\A$ is generated with $\A = \B\R$, where the elements in $\B \in\mathbb{R}^{d\times r}$ are i.i.d. $U(-\sqrt{3},\sqrt{3})$, and $\R \in \mathbb{R}^{r\times r}$ is diagonal with the $j$-th diagonal element being $d^{-\zeta_{j}}$, $0\leq \zeta_{j}\leq 0.5$. Pervasive (strong) factors have $\zeta_{j}=0$, while weak factors have $0<\zeta_{j}\leq 0.5$. In this way, the constant $C$ in (\ref{eqn:truestrength}) will be close to 1, so that $\tilde{\alpha}_{j} \approx \alpha_{j}$ and $\|\a_{j}\| \approx d_k^{\alpha_{j}}$ for $j \in [r]$. For matrix time series, we independently generate $\A_1$ and $\A_2$ using the same procedure described above. The factor loading matrix $\A_k$ for $k = 1, 2$ is generated independently with $\A_k = \B_k\R_k$, where the elements in $\B_k \in\mathbb{R}^{d_k\times r_k}$ are i.i.d. $U(-\sqrt{3},\sqrt{3})$, and $\R_k \in \mathbb{R}^{r_k\times r_k}$ is diagonal with the $j$-th diagonal element being $d_k^{-\zeta_{k,j}}$, $0\leq \zeta_{k,j}\leq 0.5$. Pervasive (strong) factors have $\zeta_{k,j}=0$, while weak factors have $0<\zeta_{k,j}\leq 0.5$.

The elements in $\f_t$ for vector time series (or $\F_t$ for matrix time series) are independent standardized AR(1) with AR coefficients 0.8. The elements in $\bepsilon_t$ (or $\E_t$) are generated based on Assumption (E1) and (E2) in \cite{Chen_Lam_2024}, where they decompose the noise series into a common component part and an independent noise part, facilitating weak cross-correlations and serial dependence in the noise series. We use the same parameters to generate $\bepsilon_t$ (or $\E_t$) as in \cite{Chen_Lam_2024}, except that we add an additional step to normalize the magnitude of the noise based on the signal-to-noise ratio $\delta$. This ratio is defined as the average ratio of standard errors of $\f_t$ and $\bepsilon_t$ (or $\F_t$ and $\E_t$), which ensures that $\frac{1}{d} \sum_{j=1}^{d} \text{var}(\e_{t,j}) = \frac{1}{\delta^2}$ (or $\frac{1}{d_1d_2} \sum_{i=1}^{d_1}\sum_{j=1}^{d_2} \text{var}(\e_{t,i,j}) = \frac{1}{\delta^2}$). We assume $\delta = 2$ for all simulation experiments in this section.

We set $r = 2$ for vector time series, and $r_1 = r_2 = 2$ for matrix time series. We consider two settings of factor strengths for vector time series:
\begin{itemize}
\item[(I)]  One strong factor and one weak factor with $\zeta_{1}=0$ and $\zeta_{2}=0.2$, so that $\alpha_{1} = 1$, $\alpha_{2} = 0.6$.

\item[(II)] Two weak factors with $\zeta_{1}=0.1$ and $\zeta_{2}=0.2$, so that $\alpha_{1} = 0.8$, $\alpha_{2} = 0.6$.
\end{itemize}
Similarly, two settings of factor strengths are considered for matrix time series:
\begin{itemize}
\item[(I)]  One strong factor and one weak factor with $\zeta_{k,1}=0$ and $\zeta_{k,2}=0.2$ for $k = 1, 2$, so that $\alpha_{k,1} = 1$, $\alpha_{k,2} = 0.6$.

\item[(II)] Two weak factors with $\zeta_{k,1}=0.1$ and $\zeta_{k,2}=0.2$ for $k = 1, 2$, so that $\alpha_{k,1} = 0.8$, $\alpha_{k,2} = 0.6$.
\end{itemize}
Each experiment is repeated for 500 times.

\subsection{Results}

For vector factor models, we consider all combinations of dimensions $d = 50, 100, 200, 400, 800$ and $T = 50, 100, 200, 400, 800$ for each of the two settings of factor strengths outlined in Section \ref{subsec:simulation_setting_chapter5}. We estimate $\hat\alpha_1$ and $\hat\alpha_2$ following the process described in Section \ref{sec:vector_factor_strength}, where $\hat\Q$ is estimated using PCA of the sample covariance matrix \citep{BaiNg2002}. Tables \ref{table:vectorSettingI} and \ref{table:vectorSettingII} record the mean and standard deviation over 500 repetitions of factor strengths estimations under different settings and dimensions.

Based on the results presented in Table \ref{table:vectorSettingI} and \ref{table:vectorSettingII}, our factor strengths estimators demonstrate good performance across all settings in vector factor models. Both $\hat\alpha_1$ and $\hat\alpha_2$ converge to the true factor strengths $\alpha_1$ and $\alpha_2$, with a particularly notable improvement as $T$ increases. Furthermore, the standard deviation of the estimators decreases with the increase in $T$ or $d$. It's essential to note that the standard deviation of estimation is influenced not only by errors in the estimation procedure but also by the fact that $\tilde\alpha_j$ is not generated to be exactly $\alpha_j$ but with some small variance (i.e., the constant $C$ in (\ref{eqn:truestrength}) is not exactly 1). Nevertheless, given that $\tilde{\alpha}_{j} \approx \alpha_{j}$, the estimated $\hat\alpha_j$ still serves as a good approximation of $\alpha_j$.

\begin{table}[!ht]

\begin{center}
  \begin{tabular}{|c|c|c|c|c|c|c|}

\hline
 & $d$ & $T = 50$ & $T = 100$ & $T = 200$ & $T = 400$ & $T = 800$\\
\hline
\hline
\multirow{5}{*}{$\hat{\alpha}_{1}$} & $50$ & 1.00(0.10) &	0.99(0.07)		&1.00(0.06)	&1.00(0.05)	&1.00(0.04)	
 \\
& $100$ &0.99(0.09)		&0.98(0.06)	&0.98(0.05)		&1.00(0.04)		&1.00(0.03)	
\\
& $200$ &0.99(0.08)	&0.98(0.06)		&0.99(0.04)	&1.00(0.03)		&1.00(0.02)	
\\
& $400$ & 0.98(0.07)  &0.99(0.04)  &1.00(0.04)		&1.00(0.03)		&1.00(0.02)	
\\
& $800$ & 0.98(0.07)  &1.00(0.04)  &0.99(0.03)		&1.00(0.02)		&1.00(0.02)
\\
\hline
&&&&&&\\[-1ex]
\hline
\multirow{5}{*}{$\hat{\alpha}_{2}$}& $50$ &0.56(0.08)&0.59(0.08)	&0.59(0.06)	&0.59(0.05)&0.59(0.05)
\\
& $100$ &0.58(0.07)	&0.59(0.07)&0.59(0.05)&0.59(0.04)&0.60(0.03)
\\
&$200$&0.59(0.06)&0.60(0.05)&0.60(0.04)	&0.60(0.03)&0.60(0.02)
\\
& $400$&0.62(0.05)&0.59(0.05)&0.60(0.04)&0.60(0.03)&0.60(0.02)
\\
& $800$ &0.64(0.04)&0.61(0.04)&0.60(0.03)&0.60(0.02)	&0.60(0.02)
\\
\hline

\end{tabular}
\end{center}
\caption{The mean and standard deviation (in brackets) of the estimated factor strengths for Setting (I) under vector factor models. The true factor strengths are $\alpha_{1} = 1$, $\alpha_{2} =  0.6$.}\label{table:vectorSettingI}
\end{table}

\begin{table}[!ht]

\begin{center}
  \begin{tabular}{|c|c|c|c|c|c|c|}

\hline
 & $d$ & $T = 50$ & $T = 100$ & $T = 200$ & $T = 400$ & $T = 800$\\
\hline
\hline
\multirow{5}{*}{$\hat{\alpha}_{1}$} & $50$ & 0.82(0.09)	&0.81(0.08)		&0.82(0.05)	&0.81(0.05)	&0.81(0.04)	
 \\
& $100$ & 0.80(0.09) &0.80(0.07)  &0.80(0.05)	&0.80(0.03)		&0.80(0.03)	
\\
& $200$ & 0.79(0.07)  &0.79(0.05) &0.80(0.04)	&0.81(0.03)		&0.80(0.03)	
\\
& $400$ & 0.80(0.07)&0.80(0.05)&0.80(0.04) &0.80(0.02)	&0.80(0.02)
\\
& $800$ & 0.81(0.05)  &0.80(0.05)  &0.80(0.03)		&0.80(0.03)		&0.80(0.02)	
\\
\hline
&&&&&&\\[-1ex]
\hline
\multirow{5}{*}{$\hat{\alpha}_{2}$}& $50$ &0.56(0.09)	&0.56(0.08)	&0.59(0.06)	&0.60(0.05)&0.59(0.05)
\\
& $100$  &0.57(0.07)&0.59(0.05)&0.59(0.05)	&0.59(0.03)&0.59(0.03)
\\
&$200$&0.59(0.05) &0.59(0.05)&0.60(0.04) &0.59(0.03)	&0.60(0.02)
\\
& $400$  &0.61(0.05)  &0.60(0.05) 	&0.60(0.03)	&0.60(0.03)		&0.60(0.02)
\\
& $800$ &0.63(0.03) &0.61(0.04) &0.60(0.03) &0.60(0.02) &0.60(0.02)
\\
\hline

\end{tabular}
\end{center}
\caption{The mean and standard deviation (in brackets) of the estimated factor strengths for Setting (II) under vector factor models. The true factor strengths are $\alpha_{1} = 0.8$, $\alpha_{2} =  0.6$.}\label{table:vectorSettingII}
\end{table}

For matrix factor models, we consider the following five settings of different dimensions for $d_1$ and $d_2$:
\begin{itemize}
  \item[i.] $d_1=d_2=25$; \;\;\;
  \item[ii.] $d_1=d_2=50$; \;\;\;
  \item[iii.] $d_1=d_2=100$;
  \item[iv.] $d_1 = 25$, \; $d_2 = 50$;
  \item[v.] $d_1 = 50$, \; $d_2 = 100$.
\end{itemize}
We consider all combinations of $T = 50, 100, 200, 400, 800$, and the above five settings of dimensions for each of the two settings of factor strengths outlined in Section \ref{subsec:simulation_setting_chapter5}. We estimate $\hat{\alpha}_{1,1}$, $\hat{\alpha}_{1,2}$, $\hat{\alpha}_{2,1}$, and $\hat{\alpha}_{2,2}$ following the process described in Section \ref{sec:matrix_factor_strength}, where $\hat\Q_1$ and $\hat\Q_2$ are estimated using the pre-averaging and iterative projection algorithm developed in \cite{Chen_Lam_2024}. Table \ref{table:matrixSettingI} and \ref{table:matrixSettingII} record the mean and standard deviation over 100 repetitions of factor strengths estimations under different settings and dimensions.

From Table \ref{table:matrixSettingI} and \ref{table:matrixSettingII}, our estimation procedure performs effectively across all settings in matrix factor models. The identifiability condition (\ref{eqn:new_identifiability_condition}) efficiently allocates factor strengths between $\A_1$ and $\A_2$. When $d_1 = d_2$, we estimate relatively similar factor strengths for $\A_1$ and $\A_2$ since they are indistinguishable. Moreover, all estimated factor strengths converge to the true values as $T$ and $d$ increase. In cases where $d_1 \neq d_2$, the estimated factor strengths on $\A_1$ and $\A_2$ are allocated based on the relative magnitudes of $d_1$ and $d_2$, contributing to the recovery of true factor strengths. This tendency is particularly pronounced in Setting (I), where the strongest factors on $\A_1$ and $\A_2$ are pervasive.


\begin{table}[!ht]

\begin{center}
  \begin{tabular}{|c|c|c|c|c|c|c|}

\hline
 & $(d_1,d_2)$ & $T = 50$ & $T = 100$ & $T = 200$ & $T = 400$ & $T = 800$\\
\hline
\hline
\multirow{5}{*}{$\hat{\alpha}_{1,1}$} & $(25,25)$ & 1.00(0.07) 	& 1.00(0.06)  &	0.99(0.05)	&0.99(0.05)	&1.00(0.04)
 \\
& $(50,50)$&  0.99(0.06)  & 1.00(0.05) 	& 1.00(0.04)&	1.00(0.03)		&1.00(0.03)	
\\
& $(100,100)$ & 0.99(0.04) &	0.99(0.03)	&	1.00(0.02)&	1.00(0.02)		&1.00(0.02)\\
&&&&&&\\[-5pt]
& $(25,50)$& 1.01(0.08) 	&0.99(0.06)&0.99(0.05) &0.99(0.04) &0.99(0.04)

\\
& $(50,100)$& 0.98(0.06)  &1.00(0.04) &	0.99(0.04) &0.99(0.03)	&	0.99(0.02)
\\
\hline
&&&&&&\\[-1ex]
\hline
\multirow{5}{*}{$\hat{\alpha}_{1,2}$}& $(25,25)$ & 0.55(0.11) &0.57(0.09) &	0.58(0.09) & 0.58(0.06) & 0.59(0.06)\\
& $(50,50)$ & 0.57(0.08) & 0.58(0.07) & 0.59(0.07) &0.59(0.05) & 0.60(0.04)\\
&$(100,100)$ &0.57(0.07) &0.59(0.06)&	0.59(0.04) &0.59(0.04) &	0.59(0.03)\\
&&&&&&\\[-5pt]
& $(25,50)$&0.54(0.11)&0.55(0.10) &0.57(0.08) &0.57(0.06) &0.57(0.06)
\\
& $(50,100)$ &0.57(0.09)&0.56(0.07)&0.59(0.06)	&0.59(0.05)&0.59(0.04)

\\
\hline
&&&&&&\\[-1ex]
\hline
\multirow{5}{*}{$\hat{\alpha}_{2,1}$}& $(25,25)$ & 1.00(0.07)  &1.00(0.05) & 0.99(0.05)&	0.99(0.05)	&1.00(0.04)
\\
& $(50,50)$ & 0.99(0.06) &0.99(0.05)& 1.00(0.04)&1.00(0.03)&1.00(0.03)
\\
&$(100,100)$ & 0.99(0.04)	&0.99(0.03) &1.00(0.02)		&1.00(0.02) 	&1.00(0.02)	
\\
&&&&&&\\[-5pt]
& $(25,50)$& 1.01(0.06) &1.00(0.04)&	1.00(0.04)&	1.00(0.03)&	1.00(0.03)
\\
& $(50,100)$ & 0.99(0.05)  &1.01(0.03) &	1.00(0.03) &1.00(0.02)  &1.00(0.02)

\\
\hline

&&&&&&\\[-1ex]
\hline
\multirow{5}{*}{$\hat{\alpha}_{2,2}$}& $(25,25)$ &	0.55(0.12)&0.56(0.09) &0.58(0.08)&	0.58(0.06)	&0.58(0.06) \\
& $(50,50)$  &0.59(0.09) &0.59(0.07) &0.58(0.06)  &0.59(0.04)  &0.59(0.04)
\\
&$(100,100)$  &0.58(0.08)&0.58(0.06)&0.59(0.04)&0.59(0.04)&0.59(0.03)
\\
&&&&&&\\[-5pt]
& $(25,50)$&0.58(0.10)&	0.59(0.07)&	0.59(0.06)&	0.59(0.04)&	0.60(0.05)
\\
& $(50,100)$ &0.59(0.08)&0.58(0.06)&0.59(0.05) &0.60(0.04)&0.60(0.03)
\\
\hline

\end{tabular}
\end{center}
\caption{The mean and standard deviation (in brackets) of the estimated factor strengths for Setting (I) under matrix factor models. The true factor strengths are $\alpha_{1,1} = \alpha_{2,1}= 1$, $\alpha_{1,2} = \alpha_{2,2}= 0.6$.}\label{table:matrixSettingI}
\end{table}

\begin{table}[!ht]

\begin{center}
  \begin{tabular}{|c|c|c|c|c|c|c|}

\hline
 & $(d_1,d_2)$ & $T = 50$ & $T = 100$ & $T = 200$ & $T = 400$ & $T = 800$\\
\hline
\hline
\multirow{5}{*}{$\hat{\alpha}_{1,1}$} & $(25,25)$ & 0.82(0.07)	&0.81(0.05)  &0.80(0.05)& 0.81(0.05)  &0.81(0.05)

 \\
& $(50,50)$ & 0.80(0.06) &0.81(0.04)&0.80(0.03)	&0.80(0.03)		&0.81(0.03)	
\\
& $(100,100)$ & 0.79(0.05)  &0.80(0.03)	&0.80(0.02)&0.80(0.02)		&0.80(0.02)
\\
&&&&&&\\[-5pt]
& $(25,50)$ & 0.79(0.07) &0.79(0.05) & 0.78(0.05)	&0.78(0.05)		&0.78(0.04)	
\\
& $(50,100)$ & 0.78(0.06)  &0.78(0.05) & 0.78(0.03)  &0.78(0.03)  &0.78(0.02)

\\
\hline
&&&&&&\\[-1ex]
\hline
\multirow{5}{*}{$\hat{\alpha}_{1,2}$}& $(25,25)$ & 0.56(0.10)& 0.55(0.08) &0.58(0.06)&0.58(0.06)&0.58(0.06)
\\
& $(50,50)$  &0.57(0.06) &0.58(0.06) &0.58(0.05)	&0.59(0.04)&0.59(0.04)
\\
&$(100,100)$ &0.58(0.08)&0.60(0.05)	&0.59(0.03)	&0.60(0.03)	&0.59(0.02)
\\
&&&&&&\\[-5pt]
& $(25,50)$ &0.53(0.10)&0.54(0.08)& 0.55(0.07)&0.55(0.06)&0.55(0.06)
\\
& $(50,100)$ &0.56(0.08)&0.56(0.06)&0.57(0.04)&0.57(0.04)&0.57(0.04)

\\
\hline
&&&&&&\\[-1ex]
\hline
\multirow{5}{*}{$\hat{\alpha}_{2,1}$}& $(25,25)$ & 0.83(0.07) &0.80(0.06) &0.81(0.05)  &0.81(0.04)	 &0.81(0.05)	
\\
& $(50,50)$ & 0.81(0.05)  &0.81(0.04) &0.80(0.04)&0.81(0.03)  &0.81(0.03)
\\
&$(100,100)$ & 0.79(0.04) &0.80(0.03) &0.80(0.03)	&0.80(0.02)		&0.80(0.02)	
\\
&&&&&&\\[-5pt]
& $(25,50)$ & 0.83(0.06)  &0.83(0.04)  &0.81(0.04)	&0.82(0.03)		&0.82(0.03)	
\\
& $(50,100)$ & 0.82(0.04)  &0.82(0.04) &0.82(0.03)		&0.82(0.02)		&0.82(0.02)	
\\
\hline

&&&&&&\\[-1ex]
\hline
\multirow{5}{*}{$\hat{\alpha}_{2,2}$}& $(25,25)$ & 0.55(0.09) &0.57(0.08)&0.57(0.06)&0.57(0.06)&0.58(0.06)
\\
& $(50,50)$ &0.57(0.06)&0.57(0.06) &0.59(0.05) &0.59(0.04)&0.59(0.04)
\\
&$(100,100)$   &0.60(0.06)&0.59(0.05)&0.60(0.04)&0.60(0.03)	&0.59(0.02)
\\
&&&&&&\\[-5pt]
& $(25,50)$&0.58(0.08)&0.60(0.06) &0.62(0.05)&0.61(0.04)&0.61(0.04)
\\
& $(50,100)$ &0.60(0.06) &0.61(0.05)&0.61(0.03)&0.62(0.03)&0.62(0.02)
\\
\hline

\end{tabular}
\end{center}
\caption{The mean and standard deviation (in brackets) of the estimated factor strengths for Setting (II) under matrix factor models. The true factor strengths are $\alpha_{1,1} = \alpha_{2,1}= 0.8$, $\alpha_{1,2} = \alpha_{2,2}= 0.6$.}\label{table:matrixSettingII}
\end{table}

{

\subsection{NYC taxi traffic analysis}\label{subsec:realdata}

We analyze taxi traffic pattern in New York city. The data includes all individual taxi rides operated by Yellow Taxi within New York City, published at

\href{https://www1.nyc.gov/site/tlc/about/tlc-trip-record-data.page}
{https://www1.nyc.gov/site/tlc/about/tlc-trip-record-data.page}.

To simplify the discussion, we only consider rides within Manhattan Island. The dataset contains 1.1 billion trip records within the period of January 1, 2011 to December 31,
2021. Each trip record includes fields capturing pick-up and drop-off dates/times, pick-up and drop-off locations, trip distances, itemized fares, rate types, payment types, and driver-reported passenger counts. Our study focuses on the pick-up and drop-off dates/times, and pick-up and drop-off locations of each ride.

The pick-up and drop-off locations in Manhattan are coded according to 69 predefined zones and we will use them to classify the pick-up and drop-off locations. While \cite{Chen_Lam_2024} and \cite{Chenetal2018} further divide each day into 24 hourly periods and analyze the $\cX_t \in \mathbb{R}^{69\times69\times24}$ tensor time series to estimate factor loadings and the number of factors, our focus in this paper is on estimating factor strengths in matrix time series. To achieve this, we record the total number of rides moving among the zones within specific hours on each day.  As an example, we utilize data from 10pm to 12am on each non-business day, representing taxi traffic patterns during nighttime \citep{Chen_Lam_2024, Chenetal2022}. We analyze the non-business-day series within the period from January 1, 2011, to December 31, 2021, encompassing 1248 days. Thus, $\X_t \in \mathbb{R}^{69\times69}$ for each day, where $x_{i_1,i_2,t}$ represents the number of trips from zone $i_1$ (the pick-up zone) to zone $i_2$ (the drop-off zone) between 10pm and 12am on day $t$. Similar analysis can be conducted to examine traffic patterns by aggregating data from other hours as well.

\cite{Chen_Lam_2024} utilizes their bootstrapped correlation thresholding method to estimate the number of factors for both pick-up and drop-off locations, resulting in $\hat{r}_1 = \hat{r}_2 = 3$. This observation suggests the potential existence of weak factors, as opposed to other rank estimators designed to analyze only pervasive factors, all of which yield $\hat{r}_1 = \hat{r}_2 = 1$ \citep{Yuetal2022a, Heetal2022a, Heetal2022b, Heetal2021}. To assess the factor strengths for these potential weak factors, we incorporate $\hat{r}_1 = \hat{r}_2 = 3$ in our analysis.

To estimate factor strengths, we first obtain estimators of $\hat\Q_1$ and $\hat\Q_2$ using the pre-averaging and iterative projection method proposed by \cite{Chen_Lam_2024}. This method is specifically designed to provide more accurate estimators of the factor loadings particularly in the presence of weak factors. Subsequently, using the estimated $\hat\Q_1$ and $\hat\Q_2$, we calculate the estimated factor strengths using the method outlined in Section \ref{sec:matrix_factor_strength}. The heatmaps of the loading matrices $\A_1$ (for pick-up locations) and $\A_2$ (for drop-off locations) are shown in Figure \ref{fig:1} and \ref{fig:2}, respectively. The estimated factor strengths $\hat{\alpha}_{k,j}$ are summarized in Table \ref{table_nyc}.

\begin{table}[ht]
\begin{center}
  \begin{tabular}{|c|c|c|c|}

\hline
\multirow{2}{*}{Pick-up Factors} &  $\hat\alpha_{1,1}$ &  $\hat\alpha_{1,2}$ &  $\hat\alpha_{1,3}$\\
\cline{2 - 4}
 & 1.39 & 0.65 & 0.50\\
\hline
\hline
\multirow{2}{*}{Drop-off Factors}  &  $\hat\alpha_{2,1}$ &  $\hat\alpha_{2,2}$ &  $\hat\alpha_{2,3}$\\
\cline{2 - 4}
 & 1.40 & 0.82 & 0.63\\
\hline

\end{tabular}
\end{center}
\caption{Factor strengths estimators for NYC Taxi Traffic Data.}\label{table_nyc}
\end{table}

From Table \ref{table_nyc}, both $\hat\alpha_{1,1}$ and $\hat\alpha_{2,1}$ have values greater than 1, indicating that the strongest factor for both the pick-up and drop-off loadings should be pervasive with $C > 1$. This observation is further supported by Figure \ref{fig:1} and Figure \ref{fig:2}, where both figures show that the first factor has a large number of elements that are not close to zero. Specifically, both factors load most heavily on East Village, where a significant number of arts, music venues, and restaurants are located. Moreover, the values of $\hat\alpha_{1,1}$ and $\hat\alpha_{2,1}$ are very close because $d_1 = d_2$ and $r_1 = r_2$, and the identifiability condition (\ref{eqn:new_identifiability_condition}) assigns similar strengths for the strongest factors to both modes. This alignment is reasonable in practice.

Furthermore, it can be observed from Table \ref{table_nyc} that the estimated strengths of the second and third factors for both pick-up and drop-off loadings are much less than 1. This suggests that these factors are likely to be weak (non-pervasive), a conclusion supported by Figures \ref{fig:1} and \ref{fig:2} as well. In these figures, the second and third factors exhibit certain localized behavior, with many entries near zero. Specifically, Factor 2 in the pick-up loadings loads heavily on Penn Station, a major transportation hub, while Factor 2 in the drop-off loadings loads heavily on the Lower East Side, known for its nightlife and entertainment venues. Furthermore, both Factor 3 in the pick-up and drop-off loadings highlight Times Square/Theatre District, another popular tourist destination and hub of nightlife activity. It is also important to note that sparsity alone may not fully account for the presence of weak factors. For example, Factor 2 in the pick-up loadings demonstrates greater sparsity compared to Factor 3, yet it exhibits stronger estimated strengths. This suggests the potential weak influence of certain factors on some or all observed variables.


\begin{center}

\begin{figure}[!htp]
    \hspace{-0.18in}
	\includegraphics[width=18cm]{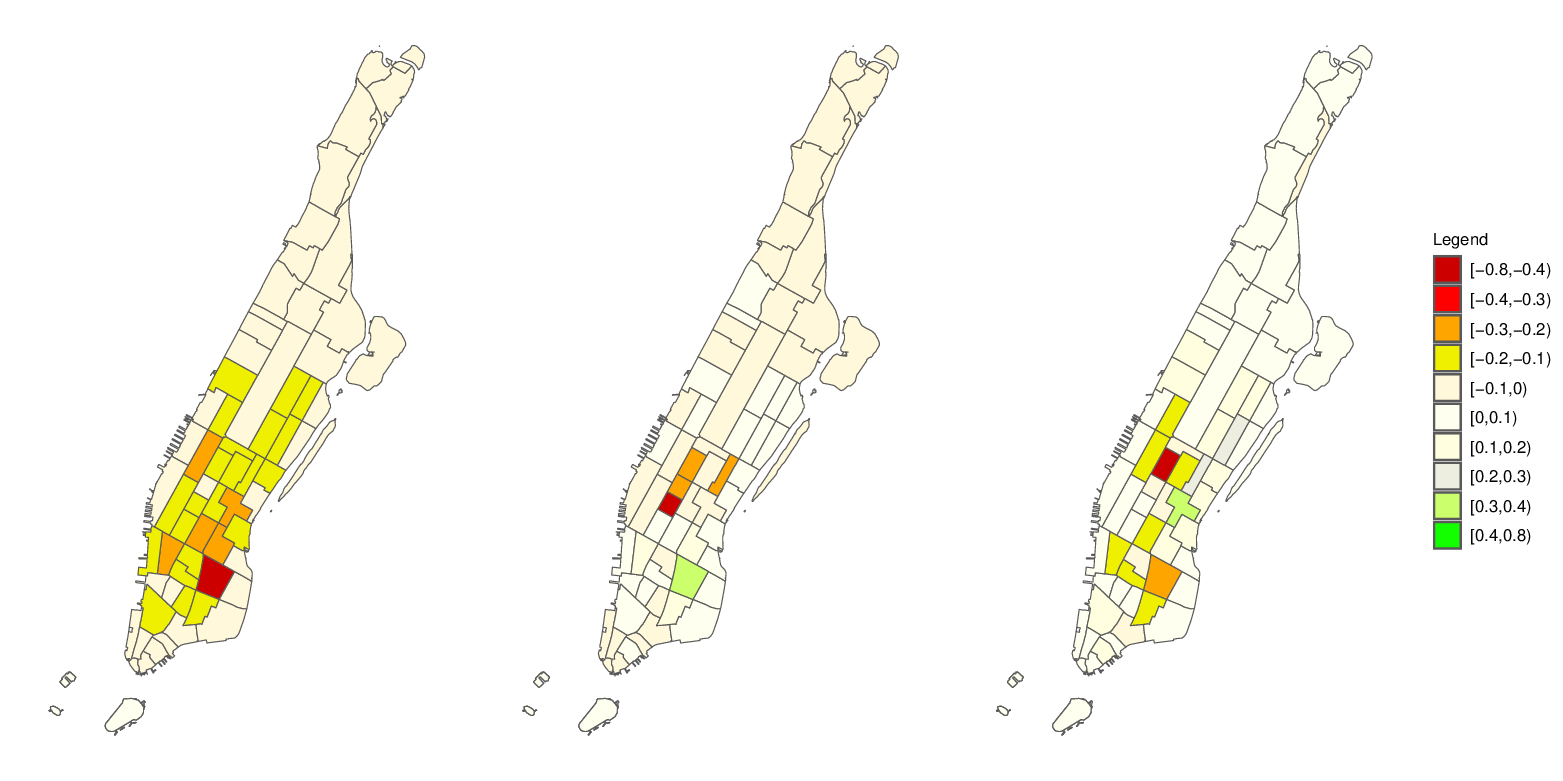}\\
	\caption{Loadings on three pickup factors.}
	\label{fig:1}
\end{figure}

\begin{figure}[!htp]
    \hspace{-0.18in}
	\includegraphics[width=18cm]{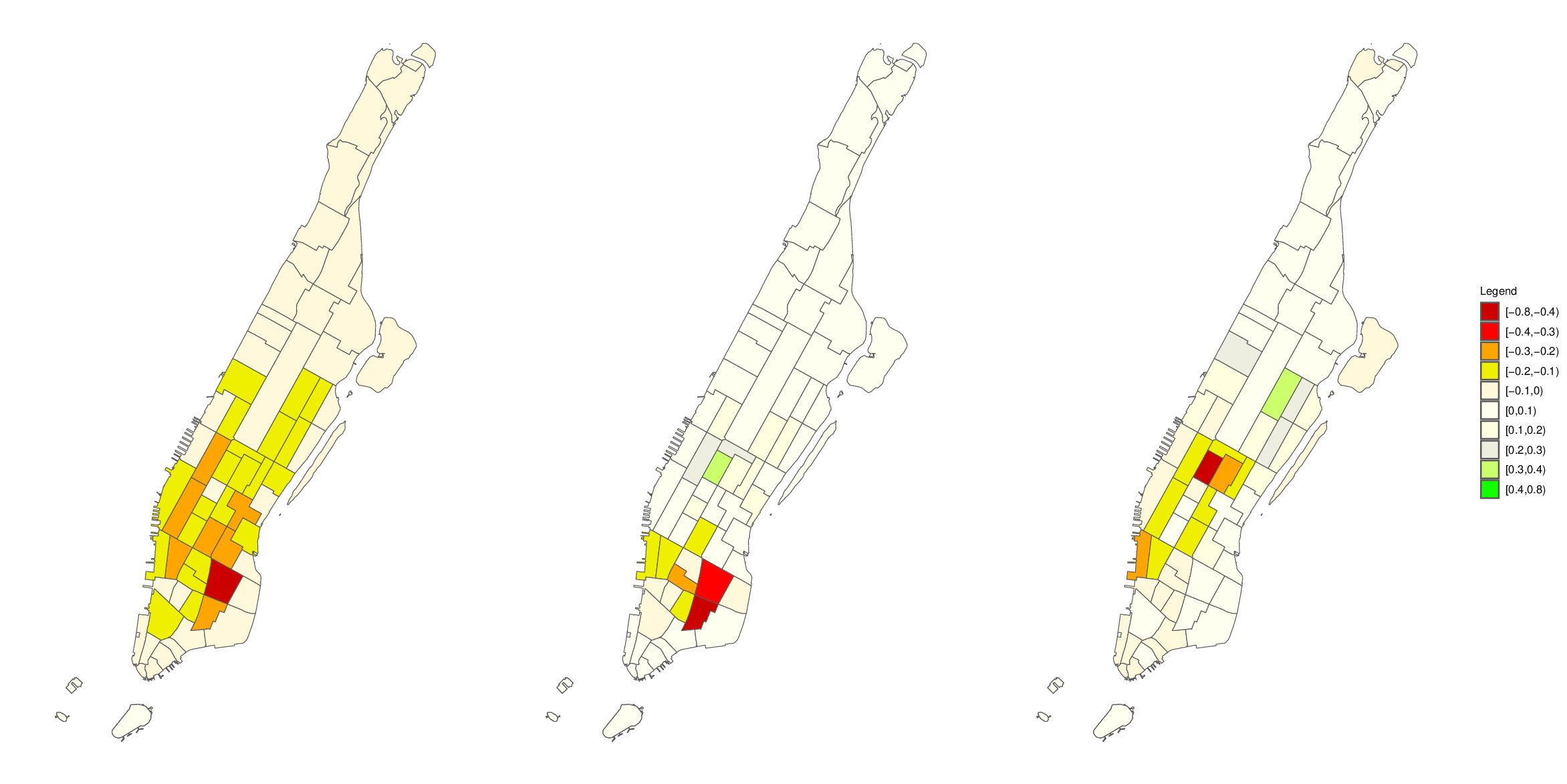}\\
	\caption{Loadings on three dropoff factors.}
	\label{fig:2}
\end{figure}

\end{center}

}

\newpage

\bibliographystyle{apalike}
\bibliography{references}

\newpage

\section{Appendix}\label{sec:appendix}

{\em Proof of Theorem \ref{thm:vector_consistency}}.
To start with, note that  (\ref{eqn:S_decom}) can be written as
\begin{align*}
    \hat{\S} = \S_{11} + \S_{12} + \S_{21} + \S_{22},
\end{align*}
where
\begin{align*}
    \S_{11}: &= \hat{\Q}^\T \Q\D^{1/2}  \left( \frac{\F^\T \F}{T} \right) \D^{1/2}\Q^\T \hat{\Q}, \\
    \S_{12}: &= \hat{\Q}^\T\Q\D^{1/2} \left( \frac{\F^\T \E}{T} \right) \hat{\Q}, \\
    \S_{21}: &= \hat{\Q}^\T \left( \frac{\E^\T \F}{T}  \right) \D^{1/2}\Q^\T\hat{\Q}, \\
    \S_{22}: &= \hat{\Q}^\T\left( \frac{\E^\T \E}{T} \right) \hat{\Q}.
\end{align*}

For each $j = 1, \cdots, r$, we have $\hat\alpha_j = \frac{\log(\hat{d}_{jj})}{\log(d)} = \frac{\log(\hat{s}_{jj})}{\log(d)}$. Thus, we need to bound the distance between $\hat{s}_{jj}$ and $d^\alpha_j$. We start with $\S_{11}$ which contains the true signal part. Denote $\M := \hat\Q^\T \Q$, then we can further decompose $\S_{11}$ as
\begin{align*}
    \S_{11} = \S_{11,1} + \S_{11,2} + \S_{11,3} + \S_{11,4},
\end{align*}
where
\begin{align*}
    \S_{11,1}: &= \D^{1/2}  \left( \frac{\F^\T \F}{T} \right) \D^{1/2}, \\
    \S_{11,2}: &= \M \D^{1/2}  \left( \frac{\F^\T \F}{T} \right) \D^{1/2} (\M - \I_r)^\T, \\
    \S_{11,3}: &= (\M - \I_r) \D^{1/2}  \left( \frac{\F^\T \F}{T} \right) \D^{1/2} \M^\T, \\
    \S_{11,4}: &= (\M - \I_r) \D^{1/2}  \left( \frac{\F^\T \F}{T} \right) \D^{1/2} (\I_r - \M)^\T.
\end{align*}

By Assumption (V2), the diagonal entries of $\frac{\F^\T \F}{T}$ are all bounded with constant magnitude $O(1)$. Thus, $(\S_{11,1})_{jj} \asymp d^\alpha_j$, which incorporates the true factor strengths. The estimation errors between $\hat{s}_{jj}$ and $d^\alpha_j$ comes from the diagonal entries of all remaining terms $\S_{11,2}, \S_{11,3}, \S_{11,4}, \S_{12}, \S_{21}, \S_{22}$, and we bound each of them accordingly.

Note that $\S_{11,2} + \S_{11,3}$ is a symmetric matrix, so by the Schur's majorization theorem, $\max_j|(\S_{11,2} + \S_{11,3})_{jj}| \leq |\lambda_1(\S_{11,2} + \S_{11,3})| \leq \norm{\S_{11,2} + \S_{11,3}}$, and
\begin{align*}
    \norm{\S_{11,2} + \S_{11,3}} &\leq 2 \norm{\M} \norm{\D} \left\|\frac{\F^\T \F}{T}\right\| \norm{\M - \I_r} \\
    &\preceq d^{\alpha_1}\norm{\M - \I_r} \\
    & =  O_{\mathbb{P}}(d^{\alpha_r}),
\end{align*}
where the last step follows as $\norm{\M - \I_r} = \norm{\hat\Q^\T(\Q - \hat\Q)} \leq \norm{\hat\Q - \Q} = O_{\mathbb{P}}(d^{\alpha_r - \alpha_1})$.
Thus, $\max_j|(\S_{11,2} + \S_{11,3})_{jj}| =  O_{\mathbb{P}}(d^{\alpha_r})$. Next,
\begin{align*}
    \norm{\S_{11,4}} &\leq \norm{\D} \left\|\frac{\F^\T \F}{T}\right\| \norm{\M - \I_r}^2 \\
    &\preceq d^{\alpha_1}\norm{\M - \I_r}^2 \\
    & =  O_{\mathbb{P}}(d^{2\alpha_r - \alpha_1}) \\
    &\preceq O_{\mathbb{P}}(d^{\alpha_r}),
\end{align*}
and $\max_j|(\S_{11,4})_{jj}| \leq |\lambda_1(\S_{11,4})| \leq \norm{\S_{11,4}} = O_{\mathbb{P}}(d^{\alpha_r})$ since $\S_{11,4}$ is also symmetric.
Similar steps can be applied to deal with $\S_{12} + \S_{21}$ and $\S_{22}$, respectively, since they are both symmetric matrices. We have
\begin{align*}
    \norm{\S_{12} + \S_{21}} &\leq 2 \norm{\D^{1/2}} \left\|\frac{\F^\T \E}{T}\right\| \\
    & \leq 2 \norm{\D^{1/2}} \sqrt{\left\|\frac{\F^\T \F}{T}\right\|} \sqrt{\left\|\frac{\E^\T \E}{T}\right\|} \\
    &\preceq O_{\mathbb{P}}\left(d^{\frac{\alpha_1}{2}}\left(1 + \sqrt{\frac{d}{T}}\right)\right),
\end{align*}
and
\begin{align*}
    \norm{\S_{22}} \leq \left\|\frac{\E^\T \E}{T}\right\| = O_{\mathbb{P}}\left(1 + \frac{d}{T}\right).
\end{align*}
Thus, $\max_j|(\S_{12} + \S_{21})_{jj}| = O_{\mathbb{P}}\left(d^{\frac{\alpha_1}{2}}\left(1 + \sqrt{\frac{d}{T}}\right)\right)$ and  $\max_j|(\S_{22})_{jj}| = O_{\mathbb{P}}\left(1 + \frac{d}{T}\right)$. Finally, we have
\begin{align}
    |\hat\alpha_j - \alpha_j| &= \left| \frac{\log(\hat{s}_{jj}) - \log(d^{\alpha_j})}{\log(d)}\right| \notag\\
    &= \left| \log\left(\frac{(\S_{11,1})_{jj} + (\S_{11,2} + \S_{11,3})_{jj} + (\S_{11,4})_{jj} + (\S_{12} + \S_{21})_{jj} + (\S_{22})_{jj}} {d^{\alpha_j}}\right)\right|\frac{1}{\log(d)} \label{eqn:rate_decomposition}\\
    &= \left| \log\left(\frac{Cd^{\alpha_j} \left[1 + (\S_{11,2} + \S_{11,3})_{jj}/d^{\alpha_j} + (\S_{11,4})_{jj}/d^{\alpha_j} + (\S_{12} + \S_{21})_{jj}/d^{\alpha_j} + (\S_{22})_{jj}/d^{\alpha_j}\right]} {d^{\alpha_j}}\right) \right|\frac{1}{\log(d)} \notag\\
    &= \frac{\log(C) + \log(R_s)}{\log(d)} \notag,
\end{align}
where $C$ is a constant, and
\begin{align*}
    R_s:= 1 + \frac{(\S_{11,2} + \S_{11,3})_{jj}}{d^{\alpha_j}} + \frac{(\S_{11,4})_{jj}}{d^{\alpha_j}} + \frac{(\S_{12} + \S_{21})_{jj}}{d^{\alpha_j}} + \frac{(\S_{22})_{jj}}{d^{\alpha_j}}.
\end{align*}
We have obtained the upper bound for each term in $\R_s$, such that
\begin{align}
    \left|\frac{(\S_{11,2} + \S_{11,3})_{jj}}{d^{\alpha_j}}\right| &= O_{\mathbb{P}}\left(d^{\alpha_r - \alpha_j} \right) \preceq O_{\mathbb{P}}(1), \label{eqn:S11,2_bound}\\
    \left|\frac{(\S_{11,4})_{jj}}{d^{\alpha_j}}\right| &= O_{\mathbb{P}}\left(d^{\alpha_r - \alpha_j} \right) \preceq O_{\mathbb{P}}(1), \label{eqn:S11,4_bound}\\
    \left|\frac{(\S_{12} + \S_{21})_{jj}}{d^{\alpha_j}}\right| &= O_{\mathbb{P}}\left((d^{\frac{\alpha_1}{2} - \alpha_j})(1 + d^{1/2}T^{-\frac{1}{2}}) \right) \preceq O_{\mathbb{P}}(1), \label{eqn:S12_bound}\\
    \left|\frac{(\S_{22})_{jj}}{d^{\alpha_j}}\right| &= O_{\mathbb{P}}\left(d^{- \alpha_j}(1 + d T^{-1}) \right) \preceq O_{\mathbb{P}}(1), \label{eqn:S22_bound}
\end{align}
where the last two lines follow from Assumption (V5). Therefore, $R_s = O_{\mathbb{P}}(1)$ and  $|\hat\alpha_j - \alpha_j| = O_{\mathbb{P}}(1/\log(d))$. This completes the proof of Theorem \ref{thm:vector_consistency}.
$\square$

{\em Proof of Theorem \ref{thm:vector_improved_rate}}. We use the  same definitions of $\S_{11,1}, \S_{11,2}, \S_{11,3}, \S_{11,4}, \S_{12}, \S_{21}, \S_{22}$ as in the proof of Theorem \ref{thm:vector_consistency}. Define $\f_j = [f_{1j},\cdots,f_{Tj}]^\T \in \mathbb{R}^T$. If $C = 1$, then
\begin{align*}
   (\S_{11,1})_{jj} &= d^{\alpha_j}\left(\mathbb{E}\left[\frac{\f_j^\T \f_j}{T} \right] + O_{\mathbb{P}}\left(\sqrt{Var\left[\frac{\f_j^\T \f_j}{T} \right]}\right) \right) \\
   &= d^{\alpha_j} + d^{\alpha_j} O_{\mathbb{P}}\left(\sqrt{\frac{\mathbb{E}\left[(\f_j^\T \f_j)^2\right]}{T^2} - 1}\right),
\end{align*}
since $\mathbb{E}[f_{tj}]^2 = 1$ for any $t \in [T], j \in [r]$ by Assumption (V2). Next, we have
\begin{align*}
    \mathbb{E}\left[(\f_j^\T \f_j)^2\right] &= \mathbb{E}\left[\sum_{t=1}^{T} f_{tj}^2 \right]^2 \\
    &= \sum_{t=1}^T \sum_{s=1}^{T} \sum_{q\geq 0} a_q^2 a_{q-|t-s|}^2 \mathbb{E}(z^4) + \sum_{t=1}^T \sum_{s=1}^{T}\left(\sum_{q \geq 0} \sum_{p \geq 0}^{p \neq q-|t-s|} a_q^2 a_p^2 \right) \mathbb{E}(z^2)^2,
\end{align*}
where $z$ is a random variable with mean 0, variance 1, and uniformly bounded fourth moment. Let $\B_{f,T} \in \mathbb{R}^{T \times T}$ to be a matrix such that $(\B_{f,T})_{ts} = \sum_{q\geq 0} a_q^2 a_{q-|t-s|}^2$. Then
\begin{align*}
    \sum_{t=1}^T \sum_{s=1}^{T} \sum_{q\geq 0} a_q^2 a_{q-|t-s|}^2 = \sum_{t=1}^T \sum_{s=1}^{T}(\B_{f,T})_{ts}
    &\leq T \| \B_{f,T} \|_1
    =T \max_t \sum_{s = 1}^T | (\B_{f,T})_{ts} | \\
    & \leq 2T \sum_{v = 0}^{T} | \sum_{q\geq 0} a_q^2 a_{q+v}^2 |
     \leq 2T \left(\sum_{q \geq 0} a_q^2\right)^2 \leq 2T.
\end{align*}
Thus, $\sum_{t=1}^T \sum_{s=1}^{T} \sum_{q\geq 0} a_q^2 a_{q-|t-s|}^2 \mathbb{E}(z^4) = O(T)$. Next, since $\mathbb{E}(z^2) = 1$,
\begin{align*}
    \sum_{t=1}^T \sum_{s=1}^{T}\left(\sum_{q \geq 0} \sum_{p \geq 0}^{p \neq q-|t-s|} a_q^2 a_p^2 \right) \mathbb{E}(z^2)^2
    \leq  \sum_{t=1}^T \sum_{s=1}^{T}\left(\sum_{q \geq 0} \sum_{p \geq 0} a_q^2 a_p^2 \right)
     \leq \sum_{t=1}^T \sum_{s=1}^{T} \left(\sum_{q \geq 0} a_q^2\right)^2
     \leq T^2.
\end{align*}
Moreover, since $\mathbb{E}\left[(\f_j^\T \f_j)^2\right] \geq T^2$ by definition, we will also have $
    \sum_{t=1}^T \sum_{s=1}^{T}\left(\sum_{q \geq 0} \sum_{p \geq 0}^{p \neq q-|t-s|} a_q^2 a_p^2 \right) \mathbb{E}(z^2)^2 \geq T^2 - O(T)
$. Thus, we finally have that
$
    \mathbb{E}\left[(\f_j^\T \f_j)^2\right] = T^2 + O(T),
$
so
\begin{align*}
    (\S_{11,1})_{jj} = d^{\alpha_j} +  d^{\alpha_j} O_{\mathbb{P}}\left(T^{-\frac{1}{2}}\right),
\end{align*}
and we can define $(\tilde\S_{11,1})_{jj}:=(\S_{11,1})_{jj} - d^{\alpha_j} = O_{\mathbb{P}}\left(d^{\alpha_j}T^{-\frac{1}{2}}\right)$.
From (\ref{eqn:rate_decomposition}), using the same notations, we have
\begin{align*}
|\hat\alpha_j - \alpha_j| &= \left| \log\left(\frac{(\S_{11,1})_{jj} + (\S_{11,2} + \S_{11,3})_{jj} + (\S_{11,4})_{jj} + (\S_{12} + \S_{21})_{jj} + (\S_{22})_{jj}} {d^{\alpha_j}}\right)\right|\frac{1}{\log(d)} \\
& = \left| \log\left(\frac{d^{\alpha_j} \left[1 + r_s\right]} {d^{\alpha_j}}\right) \right| \frac{1}{\log(d)} \\
&= \frac{\log(1 + r_s)}{\log(d)},
\end{align*}
where
\begin{align}\label{eqn:rs}
    r_s:= (\tilde\S_{11,1})_{jj}/d^{\alpha_j} + (\S_{11,2} + \S_{11,3})_{jj}/d^{\alpha_j} + (\S_{11,4})_{jj}/d^{\alpha_j} + (\S_{12} + \S_{21})_{jj}/d^{\alpha_j} + (\S_{22})_{jj}/d^{\alpha_j}.
\end{align}
If $r_s$ is small such that $r_s = o_{\mathbb{P}}(1)$, then $\log(1 + r_s) = r_s + O(r_s^2)$, so we can achieve a convergence rate such that
\begin{align}\label{eqn:rs_rate}
    |\hat\alpha_j - \alpha_j| = O(r_s/\log(d)).
\end{align}
Note that $(\tilde\S_{11,1})_{jj}/d^{\alpha_j} =  O_{\mathbb{P}}\left(T^{-1/2}\right) = o_{\mathbb{P}}(1)$ as $T \rightarrow \infty$. Also, from (\ref{eqn:S11,2_bound}) to  (\ref{eqn:S22_bound}), if Assumption (V5') is satisfied, then for any $j$ such that $\alpha_j > \alpha_r$,
\begin{align}
    \left|\frac{(\S_{11,2} + \S_{11,3})_{jj}}{d^{\alpha_j}}\right| &= O_{\mathbb{P}}\left(d^{\alpha_r - \alpha_j} \right) \preceq o_{\mathbb{P}}(1), \label{eqn:ratenew1}\\
    \left|\frac{(\S_{11,4})_{jj}}{d^{\alpha_j}}\right| &= O_{\mathbb{P}}\left(d^{\alpha_r - \alpha_j} \right) \preceq o_{\mathbb{P}}(1), \label{eqn:ratenew2}\\
    \left|\frac{(\S_{12} + \S_{21})_{jj}}{d^{\alpha_j}}\right| &= O_{\mathbb{P}}\left((d^{\frac{\alpha_1}{2} - \alpha_j})(1 + d^{1/2}T^{-\frac{1}{2}}) \right) \preceq o_{\mathbb{P}}(1), \label{eqn:ratenew3}\\
    \left|\frac{(\S_{22})_{jj}}{d^{\alpha_j}}\right| &= O_{\mathbb{P}}\left(d^{- \alpha_j}(1 + d T^{-1}) \right) \preceq o_{\mathbb{P}}(1), \label{eqn:ratenew4}
\end{align}
so we can show (\ref{eqn:rate1}) by substituting (\ref{eqn:ratenew1})(\ref{eqn:ratenew2})(\ref{eqn:ratenew3})(\ref{eqn:ratenew4}) into (\ref{eqn:rs}) and (\ref{eqn:rs_rate}). Furthermore, if Assumption (V4') is also satisfied, then for all $j \in [r]$,
\begin{align}
    \left|\frac{(\S_{11,2} + \S_{11,3})_{jj}}{d^{\alpha_j}}\right| &= o_{\mathbb{P}}\left(d^{\alpha_r - \alpha_j} \right) \preceq o_{\mathbb{P}}(1), \label{eqn:ratenew5}\\
    \left|\frac{(\S_{11,4})_{jj}}{d^{\alpha_j}}\right| &= o_{\mathbb{P}}\left(d^{\alpha_r - \alpha_j} \right) \preceq o_{\mathbb{P}}(1), \label{eqn:ratenew6}
\end{align}
and we can show (\ref{eqn:rate2}) by substituting (\ref{eqn:ratenew5}), (\ref{eqn:ratenew6}), (\ref{eqn:ratenew3}), (\ref{eqn:ratenew4}) into (\ref{eqn:rs}) and (\ref{eqn:rs_rate}). This completes the proof of Theorem \ref{thm:vector_improved_rate}.
$\square$

{

Before presenting the proof of Theorem \ref{thm:matrix_consistency}, we first state the following Lemma.

\begin{lemma}\label{lem:ratio}
    Denote $g_1 := \tr(\D_1)$, $g_2 := \tr(\D_2)$. From (\ref{eqn:tr1_approx}) and (\ref{eqn:tr2_approx}), denote $\hat{g}_1 := \left(\frac{\tr(\hat{\S}_1 + \hat\S_2)}{2} \cdot \frac{r_1 d_1}{r_2 d_2}\right)^{1/2}$ and
$\hat{g}_2 := \left(\frac{\tr(\hat{\S}_1 + \hat\S_2)}{2} \cdot \frac{r_2 d_2}{r_1 d_1}\right)^{1/2}$. Then, under Assumptions (M1) - (M5), if the identifiability condition (\ref{eqn:new_identifiability_condition}) holds, we have
    \begin{align*}
        \frac{\tr(\hat\S_1)}{g_1 g_2} = 1 + o_{\mathbb{P}}(1), \ \ \ \frac{\tr(\hat\S_2)}{g_1 g_2} = 1 + o_{\mathbb{P}}(1),
    \end{align*}
and thus,
\begin{align*}
    \frac{\hat{g}_1}{g_1} = 1 + o_{\mathbb{P}}(1), \ \ \  \frac{\hat{g}_2}{g_2} = 1 + o_{\mathbb{P}}(1),
\end{align*}
which means $\hat{g}_1$ and $\hat{g}_2$ are ratio-consistent estimates of $g_1$ and $g_2$, respectively.
\end{lemma}

{\em Proof of Lemma \ref{lem:ratio}}
We prove the results for $k = 1$ WLOG, and the results for $k = 2$ will similarly follows. Similar to the proof of Theorem \ref{thm:vector_consistency}, we start by writing (\ref{eqn:S1_decom}) as
\begin{align*}
    \hat{\S}_1 = \S_{11}^{(1)} + \S_{12}^{(1)} + \S_{21}^{(1)} + \S_{22}^{(1)},
\end{align*}
where
\begin{align*}
    \S_{11}^{(1)}: &= \hat{\Q}_1^\T \Q_1\D_1^{1/2}  \left( \frac{1}{T} \sum_{t = 1}^T \F_t \D_2 \F_t^\T \right) \D_1^{1/2}\Q_1^\T \hat{\Q}_1, \\
    \S_{12}^{(1)}: &= \hat{\Q}_1^\T\Q_1\D_1^{1/2} \left( \frac{1}{T} \sum_{t = 1}^T \F_t \D_2^{1/2} \Q_2^\T \E_t^\T \right) \hat{\Q}_1, \\
    \S_{21}^{(1)}: &= \hat{\Q}_1^\T \left( \frac{1}{T} \sum_{t = 1}^T \E_t \Q_2 \D_2^{1/2} \F_t^\T \right) \D_1^{1/2}\Q_1^\T\hat{\Q}_1, \\
    \S_{22}^{(1)}: &= \hat{\Q}_1^\T\left( \frac{1}{T} \sum_{t = 1}^T \E_t \E_t^\T \right) \hat{\Q}_1 .
\end{align*}
Let $\M_1 := \hat\Q_1^\T \Q_1$, we can further decompose $\S_{11}^{(1)}$ as
\begin{align*}
    \S_{11}^{(1)} = \S_{11,1}^{(1)} + \S_{11,2}^{(1)} + \S_{11,3}^{(1)} + \S_{11,4}^{(1)},
\end{align*}
where
\begin{align*}
    \S_{11,1}^{(1)}: &= \D_1^{1/2}  \left( \frac{1}{T} \sum_{t = 1}^T \F_t \D_2 \F_t^\T \right) \D_1^{1/2}, \\
    \S_{11,2}^{(1)}: &= \M_1 \D_1^{1/2}  \left( \frac{1}{T} \sum_{t = 1}^T \F_t \D_2 \F_t^\T \right) \D_1^{1/2} (\M_1 - \I_{r_1})^\T, \\
    \S_{11,3}^{(1)}: &= (\M_1 - \I_{r_1}) \D_1^{1/2}  \left( \frac{1}{T} \sum_{t = 1}^T \F_t \D_2 \F_t^\T \right) \D_1^{1/2} \M_1^\T, \\
    \S_{11,4}^{(1)}: &= (\M_1 - \I_{r_1}) \D_1^{1/2}  \left( \frac{1}{T} \sum_{t = 1}^T \F_t \D_2 \F_t^\T \right) \D_1^{1/2} (\I_{r_1} - \M_1)^\T.
\end{align*}
Thus,
\begin{align}\label{eqn:trace_decomposition}
    \tr(\hat\S_1) = \tr(\S_{11,1}^{(1)}) + \tr(\S_{11,2}^{(1)}) + \tr(\S_{11,3}^{(1)}) + \tr(\S_{11,4}^{(1)}) + \tr(\S_{12}^{(1)}) + \tr(\S_{21}^{(1)}) + \tr(\S_{22}^{(1)}).
\end{align}

Next, we want to show that $\frac{\tr(\hat\S_1)}{\tr(\D_1)\tr(\D_2)} = 1 + o_{\mathbb{P}}(1)$. We start by $\tr(\S_{11,1}^{(1)})$. For each $j = 1, \cdots, r_1$, we have
\begin{align}
   (\S_{11,1}^{(1)})_{jj} &= d_1^{\alpha_{1,j}} \frac{1}{T} \sum_{t=1}^T \left(\sum_{i = 1}^{r_1} d_2^{\alpha_{2,i}} f_{tji}^2 \right) \notag \\
   & = d_1^{\alpha_{1,j}} \frac{1}{T} \sum_{t=1}^T \mathbb{E}\left(\sum_{i = 1}^{r_1} d_2^{\alpha_{2,i}} f_{tji}^2 \right) + d_1^{\alpha_{1,j}} O_{\mathbb{P}}\left(\sqrt{Var\left[\frac{1}{T} \sum_{t=1}^T \left(\sum_{i = 1}^{r_2} d_2^{\alpha_{2,i}} f_{tji}^2 \right)\right]}\right). \label{eqn:S111decom}
\end{align}
For the first term in (\ref{eqn:S111decom}),
\begin{align*}
d_1^{\alpha_{1,j}} \frac{1}{T} \sum_{t=1}^T \mathbb{E}\left(\sum_{i = 1}^{r_1} d_2^{\alpha_{2,i}} f_{tji}^2 \right) = C d_1^{\alpha_{1,j}} \tr(\D_2),
\end{align*}
where $C$ is the constant such that $\|\a_{1j} \|^2 = C d_1^{\alpha_{1,j}}$. For the second term in (\ref{eqn:S111decom}), denote $\f_{ji} = [f_{1ji},\cdots,f_{Tji}]^\T \in \mathbb{R}^T$, then
\begin{align*}
    Var\left[\frac{1}{T} \sum_{t=1}^T \left(\sum_{i = 1}^{r_2} d_2^{\alpha_{2,i}} f_{tji}^2 \right)\right] &= \sum_{i = 1}^{r_2} d_2^{\alpha_{2,i}} Var\left(\frac{\sum_{t=1}^T f_{tji}^2}{T} \right) \\ & = \sum_{i = 1}^{r_2} d_2^{\alpha_{2,i}} Var\left(\frac{\sum_{t=1}^T \f_{ji}^\T \f_{ji}}{T} \right) \\
    &= \sum_{i = 1}^{r_2} d_2^{\alpha_{2,i}} O_{\mathbb{P}}(T^{-1}) \\
    &= \tr(\D_2) O_{\mathbb{P}}(T^{-1}),
\end{align*}
where the second last step follows from the proof of Theorem \ref{thm:vector_improved_rate} by simply replacing $\f_j$ in the proof of Theorem \ref{thm:vector_improved_rate} with $\f_{ji}$ here. Thus,
\begin{align}\label{eqn:signal_rate_with_C}
     (\S_{11,1}^{(1)})_{jj} =  \tr(\D_2)Cd_1^{\alpha_{1,j}} +  [\tr(\D_2)]^{1/2}d_1^{\alpha_{1,j}}O_{\mathbb{P}}(T^{-1/2}).
\end{align}
Therefore,
\begin{align*}
    \tr(\S_{11,1}^{(1)}) = \tr(\D_1)\tr(\D_2) + \tr(\D_1)[\tr(\D_2)]^{1/2}O_{\mathbb{P}}(T^{-1/2}),
\end{align*}
and the dominating term in $\tr(\S_{11,1}^{(1)})$ is $\tr(\D_1)\tr(\D_2)$. In other words,
\begin{align*}
    \frac{\tr(\S_{11,1}^{(1)})}{\tr(\D_1)\tr(\D_2)} = 1 + o_{\mathbb{P}}(1).
\end{align*}
Next, we show that $\tr(\S_{11,1}^{(1)})$ is dominating all remaining terms in (\ref{eqn:trace_decomposition}), such that all remaining terms are dominated by the rate $\tr(\D_1)\tr(\D_2) \asymp d_1^{\alpha_{1,1}}d_2^{\alpha_{2,1}}$. We have
\begin{align*}
    \tr(\S_{11,2}^{(1)}) + \tr(\S_{11,3}^{(1)}) &\leq r_1 d_1^{\alpha_{1,1}} \left\|\frac{1}{T} \sum_{t = 1}^T \F_t \D_2 \F_t^\T\right\| \norm{\M_1 - \I_{r_1}} \preceq r_1 d_1^{\alpha_{1,1}} d_2^{\alpha_{2,1}} o_{\mathbb{P}}(1) \prec d_1^{\alpha_{1,1}}d_2^{\alpha_{2,1}}, \\
    \tr(\S_{11,4}^{(1)}) &\leq r_1 d_1^{\alpha_{1,1}} \left\|\frac{1}{T} \sum_{t = 1}^T \F_t \D_2 \F_t^\T\right\| \norm{\M_1 - \I_{r_1}}^2
    \preceq r_1 d_1^{\alpha_{1,1}} d_2^{\alpha_{2,1}}o_{\mathbb{P}}(1) \prec d_1^{\alpha_{1,1}}d_2^{\alpha_{2,1}}, \\
    \tr(\S_{12}^{(1)}) + \tr(\S_{21}^{(1)}) &\leq 2 r_1 \norm{\D_1^{1/2}} \left( \frac{1}{T} \sum_{t = 1}^T \left\|\F_t\right\|  \norm{\D_2^{1/2}} \left\|\E_t\right\|\right)
    \leq r_1 O_{\mathbb{P}}\left(d_1^{\frac{\alpha_{1,1}}{2}}d_2^{\frac{\alpha_{2,1}}{2}}\sqrt{d_2 + \frac{d_1}{T}}\right)
    \prec d_1^{\alpha_{1,1}}d_2^{\alpha_{2,1}}, \\
    \tr(\S_{22}^{(1)}) &\leq r_1 \left\|\frac{1}{T} \sum_{t = 1}^T \E_t \E_t^\T\right\| = r_1O_{\mathbb{P}}\left(d_2 + \frac{d_1}{T}\right) \prec d_1^{\alpha_{1,1}}d_2^{\alpha_{2,1}},
\end{align*}
where the last two equality follows from Assumption (M5). Therefore, from (\ref{eqn:trace_decomposition}), we have
\begin{align*}
    \frac{\tr(\hat\S_1)}{\tr(\D_1)\tr(\D_2)} = 1 + o_{\mathbb{P}}(1),
\end{align*}
which proves the first part of Lemma \ref{lem:ratio} for $k = 1$. We can similarly prove the result for $k = 2$, then we will also have
\begin{align*}
    \frac{\left[\tr(\hat\S_1) + \tr(\hat\S_1)\right]/2}{\tr(\D_1)\tr(\D_2)} = 1 + o_{\mathbb{P}}(1).
\end{align*}
Thus,
\begin{align*}
    \frac{\hat{g}_1}{g_1} = \frac{\sqrt{\frac{\tr(\hat\S_1) + \tr(\hat\S_1)}{2} \frac{r_1d_1}{r_2d_2}}}{g_1} = \frac{\sqrt{g_1g_2 (1 + o_{\mathbb{P}}(1))  \frac{r_1d_1}{r_2d_2}}}{g_1} = \frac{\sqrt{g_1^2 (1 + o_{\mathbb{P}}(1))}}{g_1} = 1 + o_{\mathbb{P}}(1),
\end{align*}
where the second last step equality follows from the identifiability condition (\ref{eqn:new_identifiability_condition}) that $\frac{g_1}{r_1d_1} = \frac{g_2}{r_2d_2}$. This completes the proof of the second part of Lemma \ref{lem:ratio} for $k = 1$. Similar arguments can be applied to prove the results for $k = 2$. Thus, we complete the proof of Lemma \ref{lem:ratio}.
$\square$

{\em Proof of Theorem \ref{thm:matrix_consistency}}.
We prove the results for $k = 1$ WLOG, and the results for $k = 2$ will similarly follows. Recall the definition of $\S_{11,1}^{(1)}, \S_{11,2}^{(1)}, \S_{11,3}^{(1)}, \S_{11,4}^{(1)}, \S_{12}^{(1)}, \S_{21}^{(1)}, \S_{22}^{(1)}$ as from the proof of Lemma \ref{lem:ratio}.

For each $j = 1, \cdots, r_1$, we have $(\S_{11,1}^{(1)})_{jj}$ contains the true signal part, and from (\ref{eqn:signal_rate_with_C}), we have
\begin{align*}
     (\S_{11,1}^{(1)})_{jj} = d_1^{\alpha_{1,j}} \left( \tr(\D_2)C +  [\tr(\D_2)]^{1/2}O_{\mathbb{P}}(T^{-1/2})\right),
\end{align*}
and we can further define
\begin{align*}
    (\tilde\S_{11,1}^{(1)})_{jj}:=(\S_{11,1}^{(1)})_{jj} - \tr(\D_2) d_1^{\alpha_{1,j}} = (C - 1) \tr(\D_2) d_1^{\alpha_{1,j}} + O_{\mathbb{P}}\left(d_1^{\alpha_{1,j}}T^{-\frac{1}{2}} [\tr(\D_2)]^{1/2} \right).
\end{align*}

Next, we bound the terms $\S_{11,2}^{(1)}, \S_{11,3}^{(1)}, \S_{11,4}^{(1)}, \S_{12}^{(1)}, \S_{21}^{(1)}, \S_{22}^{(1)}$ accordingly. Similar to the proof of Theorem \ref{thm:vector_consistency}, we have
\begin{align*}
    \norm{\S_{11,2}^{(1)} + \S_{11,3}^{(1)}} &\leq 2 \norm{\M_1} \norm{\D_1} \left\|\frac{1}{T} \sum_{t = 1}^T \F_t \D_2 \F_t^\T\right\| \norm{\M_1 - \I_{r_1}} \\
    &\preceq d_1^{\alpha_{1,1}} d_2^{\alpha_{2,1}}\norm{\M_1 - \I_{r_1}} \\
    & \preceq  o_{\mathbb{P}}(d_1^{\alpha_{1,r_1}}d_2^{\alpha_{2,1}}),
\end{align*}
where the last step follows as $\norm{\M_1 - \I_{r_1}} = \norm{\hat\Q_1^\T(\Q_1 - \hat\Q_1)} \leq \norm{\hat\Q_1 - \Q_1} = o_{\mathbb{P}}(d_1^{\alpha_{1,r_1} - \alpha_{1,1}})$.
Thus, $\max_j|(\S_{11,2}^{(1)} + \S_{11,3}^{(1)})_{jj}| =  o_{\mathbb{P}}(d_1^{\alpha_{1,r_1}}d_2^{\alpha_{2,1}})$. Next,
\begin{align*}
    \norm{\S_{11,4}^{(1)}} &\leq \norm{\D_1} \left\|\frac{1}{T} \sum_{t = 1}^T \F_t \D_2 \F_t^\T\right\| \norm{\M_1 - \I_{r_1}}^2 \\
    &\preceq d_1^{\alpha_{1,1}} d_2^{\alpha_{2,1}}\norm{\M_1 - \I_{r_1}}^2 \\
    &\preceq o_{\mathbb{P}}(d_1^{\alpha_{1,r_1}} d_2^{\alpha_{2,1}}),
\end{align*}
and thus $\max_j|(\S_{11,4}^{(1)})_{jj}| = o_{\mathbb{P}}(d_1^{\alpha_{1,r_1}} d_2^{\alpha_{2,1}})$. Similarly,
\begin{align*}
    \norm{\S_{22}^{(1)}} \leq \left\|\frac{1}{T} \sum_{t = 1}^T \E_t \E_t^\T\right\| = O_{\mathbb{P}}\left(d_2 + \frac{d_1}{T}\right).
\end{align*}
and
\begin{align*}
    \norm{\S_{12}^{(1)} + \S_{21}^{(1)}} &\leq 2 \norm{\D_1^{1/2}} \left( \frac{1}{T} \sum_{t = 1}^T \left\|\F_t\right\|  \norm{\D_2^{1/2}} \left\|\E_t\right\|\right) \\
    & \leq 2 \norm{\D_1^{1/2}} \norm{\D_2^{1/2}} \sqrt{\left\|\frac{1}{T}\sum_{t=1}^T\frac{\F_t^\T \F_t}{T}\right\|} \sqrt{\left\|\frac{1}{T}\sum_{t=1}^T\frac{\E_t \E_t^\T}{T}\right\|} \\
    &\preceq O_{\mathbb{P}}\left(d_1^{\frac{\alpha_{1,1}}{2}}d_2^{\frac{\alpha_{2,1}}{2}}\sqrt{d_2 + \frac{d_1}{T}}\right),
\end{align*}
Thus, $\max_j|(\S_{12}^{(1)} + \S_{21}^{(1)})_{jj}| = O_{\mathbb{P}}\left(d_1^{\frac{\alpha_{1,1}}{2}}d_2^{\frac{\alpha_{2,1}}{2}} \left[d_2 + \frac{d_1}{T}\right]^{1/2} \right)$ and  $\max_j|(\S_{22}^{(1)})_{jj}| =  O_{\mathbb{P}}\left(d_2 + \frac{d_1}{T}\right)$.

Now, we have $\alpha_{1,j} = \log((\hat\S_1)_{jj} / \hat{g}_2) / \log(d_1)$. Thus,
\begin{align*}
    |\hat\alpha_{1,j} - \alpha_{1,j}| &= \left| \frac{\log((\hat\S_1)_{jj} / \hat{g}_2) - \log(d^{\alpha_{1,j}})}{\log(d_1)}\right| \notag\\
    &= \left| \log\left(\frac{(\S_{11,1}^{(1)})_{jj} + (\S_{11,2}^{(1)} + \S_{11,3}^{(1)})_{jj} + (\S_{11,4}^{(1)})_{jj} + (\S_{12}^{(1)} + \S_{21}^{(1)})_{jj} + (\S_{22})^{(1)}_{jj}} {\hat{g}_2 d_1^{\alpha_{1,j}}}\right)\right|\frac{1}{\log(d_1)} \\
    & = \left| \log\left(\frac{g_2 d_1^{\alpha_{1,j}} \left[1 + r_s^{(1)}\right]} {\hat{g}_2 d_1^{\alpha_{1,j}}}\right) \right| \frac{1}{\log(d_1)}
\end{align*}
where
\begin{align}\label{eqn:rs1}
    r_s^{(1)}:= \frac{(\tilde\S_{11,1}^{(1)})_{jj}}{g_2 d_1^{\alpha_{1,j}}} + \frac{(\S_{11,2}^{(1)} + \S_{11,3}^{(1)})_{jj}}{g_2 d_1^{\alpha_{1,j}}} + \frac{(\S_{11,4}^{(1)})_{jj}}{g_2 d_1^{\alpha_{1,j}}} + \frac{(\S_{12}^{(1)} + \S_{21}^{(1)})_{jj}}{g_2 d_1^{\alpha_{1,j}}} + \frac{(\S_{22}^{(1)})_{jj}}{g_2d_1^{\alpha_{1,j}}}.
\end{align}
For each term in (\ref{eqn:rs1}), since $g_2 \asymp d_2^{\alpha_{2,1}}$, we can obtain
\begin{align*}
    \frac{(\tilde\S_{11,1}^{(1)})_{jj}}{g_2 d_1^{\alpha_{1,j}}} &= O_{\mathbb{P}}(d_2^{-\frac{\alpha_{2,1}}{2}}T^{-1/2}) + C - 1 \preceq o_{\mathbb{P}}(1) + C - 1,\\
    \frac{(\S_{11,2}^{(1)} + \S_{11,3}^{(1)})_{jj}}{g_2 d_1^{\alpha_{1,j}}} &= o_{\mathbb{P}}(d_1^{\alpha_{1,r_1} - \alpha_{1,j}}) \preceq o_{\mathbb{P}}(1) \\
    \frac{(\S_{11,4}^{(1)})_{jj}}{g_2 d_1^{\alpha_{1,j}}} &= o_{\mathbb{P}}(d_1^{\alpha_{1,r_1} - \alpha_{1,j}}) \preceq o_{\mathbb{P}}(1), \\
    \frac{(\S_{12}^{(1)} + \S_{21}^{(1)})_{jj}}{g_2 d_1^{\alpha_{1,j}}} &= O_{\mathbb{P}}\left(d_1^{\frac{\alpha_{1,1}}{2} - \alpha_{1,j}} d_2^{-\frac{\alpha_{2,1}}{2}} \left[d_2 + d_1/T\right]^{1/2}\right) \preceq o_{\mathbb{P}}(1), \\
    \frac{(\S_{22}^{(1)})_{jj}}{g_2d_1^{\alpha_{1,j}}} &= O_{\mathbb{P}}\left(d_1^{-\alpha_{1,j}} d_2^{-\alpha_{2,1}}(d_2 + \frac{d_1}{T}) \right) \preceq o_{\mathbb{P}}(1),
\end{align*}
where the last two formulas follow from Assumption (M5). Finally, if $C \neq 1$, then the constant $C - 1$ dominates $r_s^{(1)}$, and thus
\begin{align*}
    |\hat\alpha_{1,j} - \alpha_{1,j}| = \left| \log\left(\frac{g_2 \left[C + o_{\mathbb{P}}(1)\right]} {\hat{g}_2}\right) \right| \log(d_1)^{-1} = O_{\mathbb{P}}\left(\frac{\log(C)}{\log(d_1)}\right) = O_{\mathbb{P}}\left(\frac{1}{\log(d_1)}\right).
\end{align*}
If $C = 1$, then we have $r_s^{(1)} = o_{\mathbb{P}}(1)$, thus $\log(1 + r_s) = r_s + O(r_s^2)$, and we have a faster convergence rate of
\begin{align*}
    |\hat\alpha_{1,j} - \alpha_{1,j}| &= O_{\mathbb{P}}\left(\frac{r_s^{(1)}}{\log(d_1)}\right) \\
    &= o_{\mathbb{P}}\left(\frac{d_1^{\alpha_{1,r_1} - \alpha_{1,j}}}{\log(d_1)} \right) + O_{\mathbb{P}}\left(\frac{d_2^{-\frac{\alpha_{2,1}}{2}}T^{-1/2} + d_1^{\frac{\alpha_{1,1}}{2} - \alpha_{1,j}} d_2^{-\frac{\alpha_{2,1}}{2}} \left[d_2 + d_1/T\right]^{1/2} + d_1^{-\alpha_{1,j}} d_2^{-\alpha_{2,1}}(d_2 + \frac{d_1}{T})}{\log(d_1)}\right).
\end{align*}
This completes the proof of Theorem \ref{thm:matrix_consistency}.
$\square$

}

\end{document}